\documentclass{article}
\usepackage[utf8]{inputenc}
\usepackage{graphicx}
\usepackage{xfrac}
\usepackage{xcolor}
\usepackage{comment}
\usepackage{hyperref}
\hypersetup{ 
backref=true, % link backreferences (i.e. citations to their list)
bookmarks=true, % show bookmarks bar?
unicode=false, % non-Latin characters in Acrobats bookmarks
pdftoolbar=true, % show Acrobats toolbar?
pdfmenubar=true, % show Acrobats menu?
pdffitwindow=false, % window fit to page when opened
pdfstartview={FitH}, % fits the width of the page to the window
pdftitle={}, % title
pdfauthor={}, % author
pdfsubject={}, % subject of the document
pdfkeywords={}{}, % list of keywords
pdfnewwindow=true, % links in new window
colorlinks=true, % false: boxed links; true: colored links
linkcolor=cyan, % colour of internal links
citecolor=brown, % colour of links to bibliography
urlcolor=blue,
}
\usepackage{float}
\usepackage{makecell}
\usepackage{authblk}
\usepackage[style=nature]{biblatex}
\title{Estimates of the social cost of carbon have increased over time}
\author[1,2,3,4,5,6]{Richard S.J. Tol}
\affil[1]{Department of Economics, Jubilee Building, University of Sussex, Falmer, BN1 9SL, United Kingdom, r.tol@sussex.ac.uk}
\affil[2]{Institute for Environmental Studies, Vrije Universiteit, Amsterdam, The Netherlands}
\affil[3]{Department of Spatial Economics, Vrije Universiteit, Amsterdam, The Netherlands}
\affil[4]{Tinbergen Institute, Amsterdam, The Netherlands}
\affil[5]{CESifo, Munich, Germany}
\affil[6]{Payne Institute for Public Policy, Colorado School of Mines, Golden, CO, USA}

\begin{document}

\maketitle

%\textbf{Preprint} (older version, different title) \href{https://ideas.repec.org/p/sus/susewp/0821.html}{IDEAS/RePEc}

%\textbf{Classification} Social sciences, economic sciences

%\textbf{Keywords} social cost of carbon, meta-analysis, climate policy

\begin{abstract}
A meta-analysis of published estimates shows that the social cost of carbon has increased as knowledge about climate change accumulates. Correcting for inflation and emission year and controlling for the discount rate, kernel density decomposition reveals a non-stationary distribution. In the last 10 years, estimates of the social cost of carbon have increased from \$33/tC to \$146/tC for a high discount rate and from \$446/tC to \$1925/tC for a low discount rate. Actual carbon prices are almost everywhere below its estimated value and should therefore go up.
\\

\begin{center}\textbf{Significance statement}\end{center}

Greenhouse gas emissions should be taxed at the social cost of carbon. Many estimates have been published, ranging from -\$1,000/tC to +\$110,000,000/tC. I use kernel density estimation to reflect the large and asymmetric uncertainty, and kernel decomposition to test for changes over time. Correcting for inflation and emission year, and controlling for the discount rate, I show that estimates of the social cost of carbon have increased over time\textemdash estimates have more than quadrupled over the last 10 years. This justifies a tightening of climate policy.
\end{abstract}

\begin{refsection}

\newpage \section*{Main}
The social cost of carbon is a key indicator of the seriousness of climate change. Have its estimates changed over time? Should we raise our ambitions to reduce greenhouse gas emissions? Have we learned since the first estimate was published in 1982\cite{Nordhaus1982}? There is broad agreement among scholars that greenhouse gas emissions should be taxed, but the uncertainty about the optimal level of that tax is very large. I estimate the probability distribution of published estimates of the social cost of carbon and how it changes over time. I develop and apply a non-parametric test for the stationarity of this, and apply a range of other statistical tests, to show that we have. An upwards trend can be discerned. Climate policy should be intensified.

The social cost of carbon is the damage done, at the margin, by emitting more carbon dioxide into the atmosphere. If evaluated along the optimal emissions trajectory, the social cost of carbon equals the Pigou tax\cite{Pigou1920, Bator1958} that internalizes the externality and restores the economy to its Pareto optimum\cite{Pareto1906} where no one can be made better off without making someone else worse off. The social cost of carbon is then the optimal carbon price. It informs the desired intensity of climate policy.

Some have argued\cite{Kaufman2020, Stern2021} that the debate on optimal climate policy is over since the Paris Agreement has set targets for international climate policy. Analysis should focus on the cheapest way of meeting these targets, and emissions should be priced based on the \emph{shadow price of carbon}, which is the scarcity value of the carbon budget.\cite{Kaufman2020} The shadow price is different from the social cost of carbon, and their growth rates are different too. However, the first stock-take of the commitments under the Paris Agreement\cite{UNFCCC2021} reveals that few countries plan to do what is needed to meet the agreed targets. The debate over the ultimate target of international climate policy, and so the debate on the social cost of carbon, is not over. Indeed, President Biden reinstated (and renamed) the Interagency Working Group on the Social Cost of Greenhouse Gases to reassess the appropriate carbon price.\cite{Biden2021}

There is a large literature on the social cost of carbon spanning four decades\cite{Nordhaus1982, Taconet2021}. The social cost of carbon depends on many things, including the total economic impact of climate change, potential tipping points, the scenarios for population, economy and emissions, changes in vulnerability and relative prices with development, the rate of degradation of carbon dioxide from the atmosphere, the rate and level of global warming, the discount rate, the distribution of impacts and inequity aversion, and the uncertainties about impacts and risk aversion. The estimates used here\textemdash 5,905 estimates in 207 papers, published before 2022\textemdash make different assumptions about all these matters.

These are estimates of the social cost of carbon for carbon dioxide emitted in the recent past. The carbon tax should increase over time (until climate change has been mitigated to the point that its marginal impacts start to fall\cite{Ploeg2014}). 94 papers estimate how fast, showing estimates of the social cost of carbon at two or more points in time, for a total of 1,974 estimates of the \textit{growth rate} of the social cost of carbon. 

I apply meta-analysis to these estimates. Meta-analysis is not the only way to make the social cost of carbon more transparent. Sensitivity analysis\cite{Anthoff2013} and model comparison\cite{Diaz2017} are insightful too. Decomposition of model updates\cite{Nordhaus2018cc, Hansel2020} helps to understand the evolution of estimates, but only within the confines of a single model. Closed-form equations \cite{Golosov2014, Bremer2021} give an exact relationship between input parameters and the social cost of carbon, but require rather restrictive and unrealistic assumptions. Only meta-analysis can show how the entire literature has evolved over time.

\begin{figure}[hp]
    \centering
    \includegraphics[width=\textwidth]{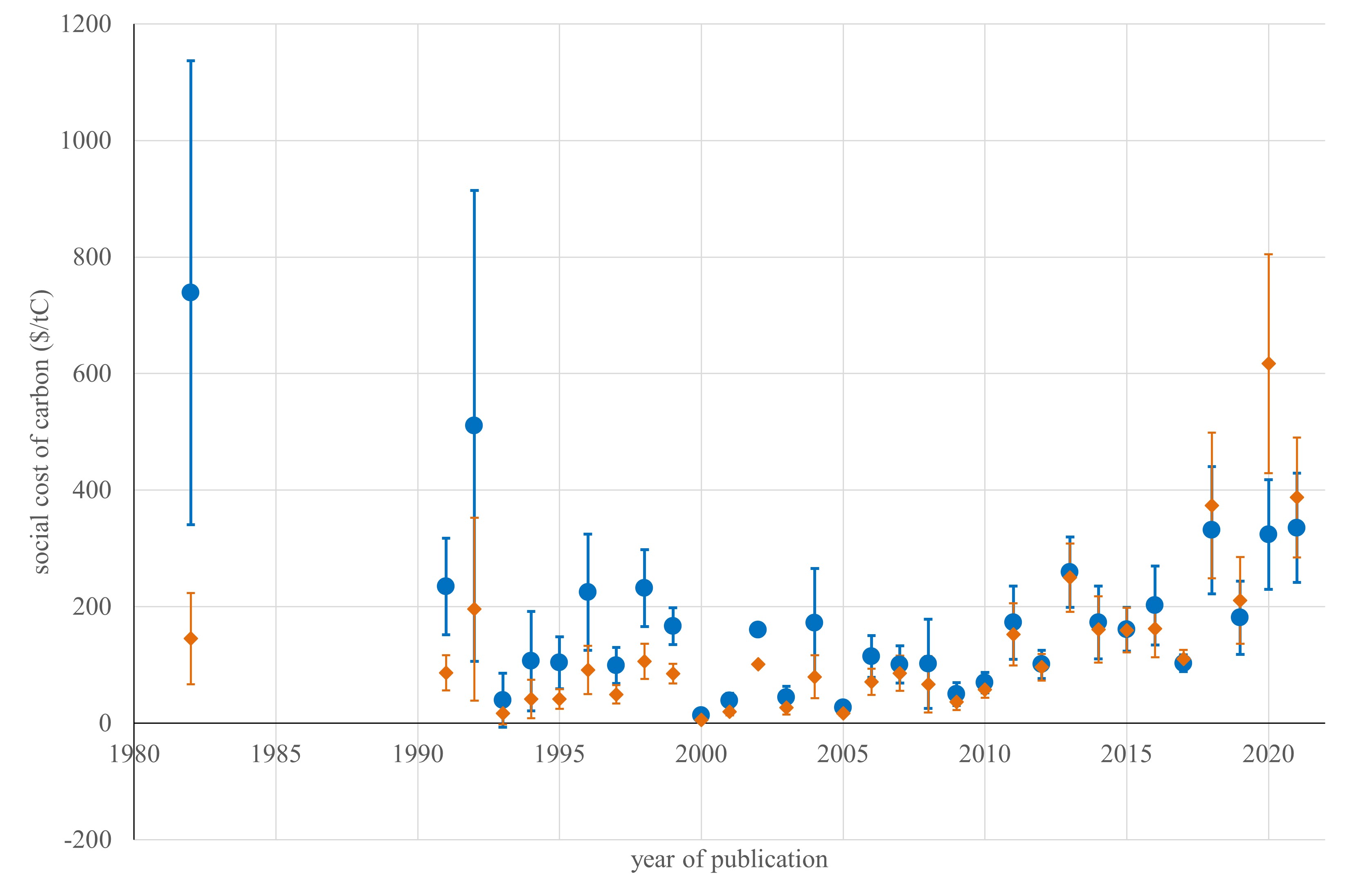}
    \caption{Average social cost of carbon by publication year. Orange diamonds are as reported, blue dots are corrected for inflation and year of emission. Error bars are plus and minus the standard deviation of the published estimates. Estimates are quality weighted and censored.}
    \label{fig:means}
\end{figure}

Figure \ref{fig:means} shows the mean and standard deviation of estimates of the social cost of carbon by year of publication. Estimates are shown with and without standardization. The social cost of carbon is expressed in 2010 US dollars per metric tonne of carbon, for emissions in the year 2010. The literature uses nominal dollars and a variety of emission years. See Figure \ref{fig:year}. Particularly, later studies report the social cost of carbon in later dollars for later emission years. Average inflation was 2.9\% over the period. The social cost of carbon grows by some 2.2\% per year; this is the average across the 94 studies that estimate its growth rate; see Figure \ref{fig:growth}. Without correcting for emission year and inflation, the \emph{apparent} trend in the social cost of carbon equals 5.1\% per year. After correction, some of the early estimates are the highest. Between 1993 and 2008, estimates went up and down without a discernible trend in either direction. Since 2009 or so, there appears to be an increase in the social cost of carbon, and three of the last four years stand out. An earlier meta-analysis finds that the social cost of carbon has not increased over time\cite{Havranek2015} but a more recent one finds that it has\cite{Wang2019}. The mean for 2021 is higher than all but two other years; paired t-tests shows the 2021 mean is statistically significantly higher than all but four other years. 

Figure \ref{fig:means} shows that the range of estimates is large and has remained large over the years, perhaps even grown recently. An assessment of the literature of the social cost of carbon should reflect that uncertainty, not just the first and second moment, but the entire probability distribution. Figure \ref{fig:means} is incomplete. In this paper, I use bespoke kernel methods to reflect the true uncertainties\textemdash including the uncertainties about parameter values, model structure, future scenarios, and ethics.

The uncertainty about the social cost of carbon is right-skewed, \cite{Tol1999} because the uncertainty about climate sensitivity is,\cite{Roe2007}, because impact functions are non-linear,\cite{Nordhaus1992} and because of risk aversion.\cite{Anthoff2009erl} This asymmetry is lost by adding and subtracting the standard error from the mean, as is done in Figure \ref{fig:means}. The uncertainty about the social cost of carbon is thick- or even fat-tailed.\cite{Weitzman2009} There is considerably more probability mass outside the Gaussian confidence bounds. The above t-tests are overconfident. If the distribution of the social cost of carbon is right-skewed and fat-tailed, then recent estimates may well be within the historical range.

Kernel densities are a flexible alternative to parametric distribution functions\cite{Takezawa2005}. Kernel densities have been used to visualize the uncertainty about the social cost of carbon.\cite{Tol2018} I here add many more observations, and decompose that uncertainty into discrete components, particularly publication periods, testing whether the components differ from one another. Simple kernel regression is helpful for specifying the relationship between two variables.\cite{Altman1992} Kernel quantile regression can be used to show this relationship across the distribution.\cite{Yu1998} However, these methods are not suitable if the explanatory variable is categorical\textemdash as is the case for assumed discount rates, authors, or recorded years of publication. The method proposed here, kernel density decomposition, works for categorical data, shows both central tendency and spread, and does not make assumptions about functional form or the shape of the probability distribution. This method, while uncommon, is therefore best suited for the problem at hand.

Kernel density decomposition offer a valid basis for statistical tests. In order to test for changes over time, I split the sample into six periods, demarcated by important events in the publication history of the social cost of carbon. These key events are the Second Assessment Report of the Intergovernmental Panel on Climate Change\cite{Pearce1996}, the Third Assessment Report of the IPCC\cite{Smith2001}, the Stern Review\cite{Stern2006}, the Obama update of the social cost of carbon\cite{IAWGSCC2013}, and the 2018 Nobel Memorial Prize in Economic Sciences.\cite{Nordhaus2019aer}

The kernel density is estimated with bespoke kernel functions, reflecting the deep and asymmetric uncertainty of the social cost of carbon. The data are weighted for quality\textemdash age, computational method, scenario, peer-review, validity, novelty\textemdash but the results are largely robust to these weights. Furthermore, implausibly high estimates are censored or, in the appendix, winsorized. The decomposition of the kernel density is based on the fact that the weighted sum of probability densities is a probability density.\cite{Quetelet1846} The statistical test is that for the equality of proportions\cite{Pearson1900}, adjusted for finite sample size. Applied to different publication periods, this is a test for the stationarity of the distribution\cite{Andrews2019} of the social cost of carbon. See \textit{Methods} for the details, the \textit{Supplementary Information} for sensitivity analyses.

\begin{figure}[hp]
    \centering
    \includegraphics[width=\textwidth]{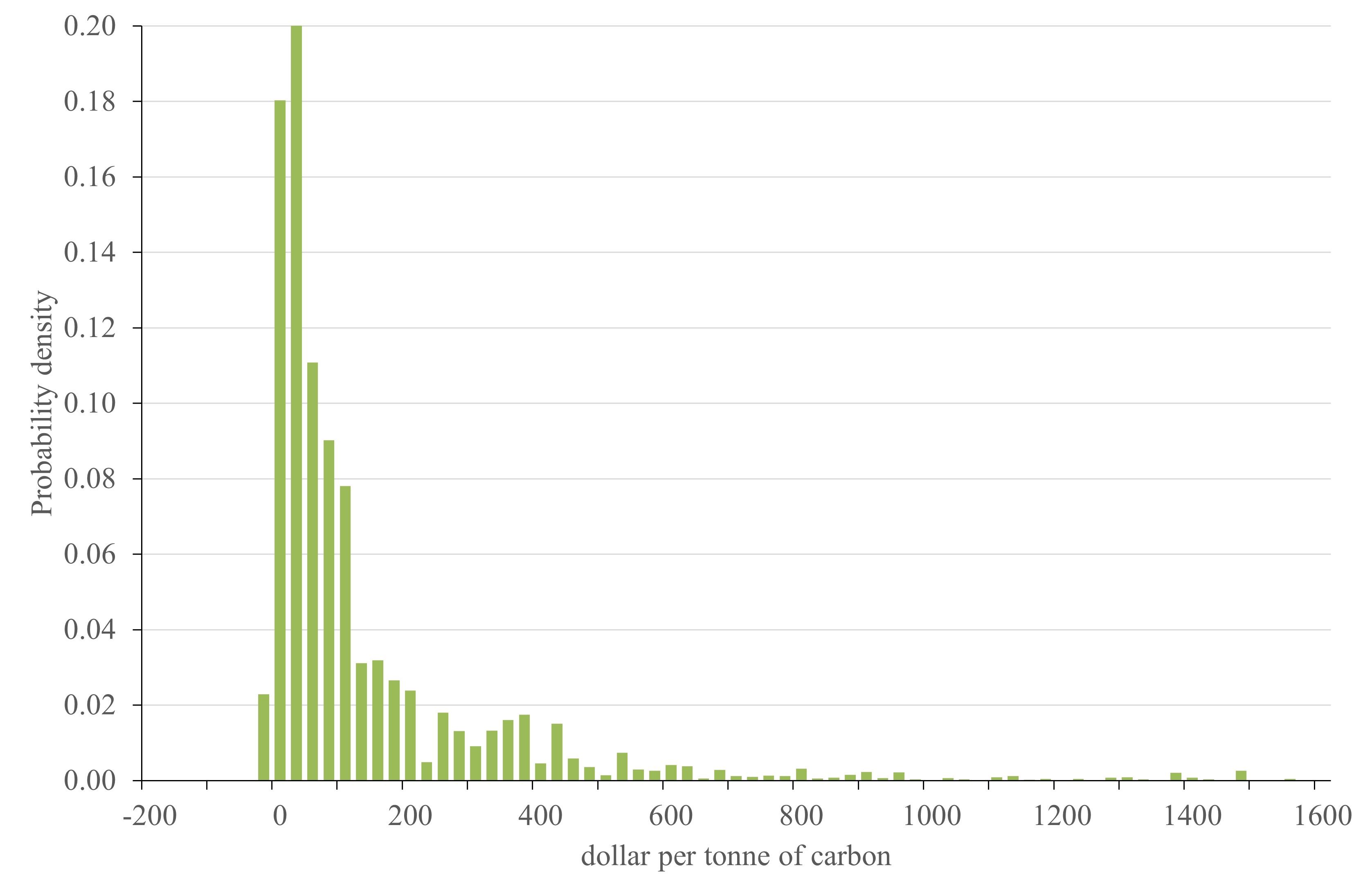}
    \includegraphics[width=\textwidth]{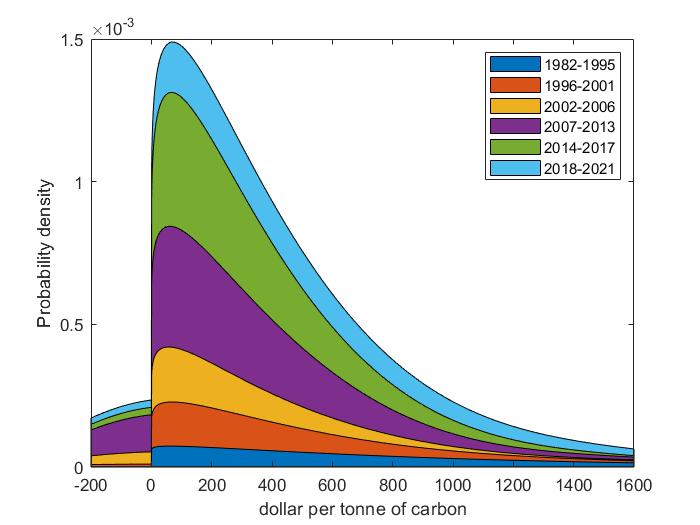}
    \caption{Histogram of the social cost of carbon. Results are quality-weighted and censored (top panel). Composite kernel density of the social cost of carbon and its composition by publication period (bottom panel).}
    \label{fig:period}
\end{figure}

Figure \ref{fig:period} (bottom panel) shows the kernel density of the social cost of carbon and its the decomposition by publication period. The kernel density has the same shape as the histogram in Figure \ref{fig:period} (top panel): There is a little probability mass below zero, a pronounced mode, and a thick right tail. Compared to the histogram, the kernel density is smooth and spread wider. This is also seen in Table \ref{tab:means}: Kernel mean and standard deviation are larger than their empirical counterparts because (1) I use the mode rather than the mean as the central estimate and (2) I assume a right-skewed kernel function.

Earlier studies exclude negative estimates, and the kernel density decomposition shows a fatter right tail for recent years. Table \ref{tab:period} confirms this. It shows the contributions of estimates of the social cost of carbon published in a particular period to the overall kernel density and its quintiles. The null hypothesis that the quintile shares are the same as the overall shares is rejected; $\chi^2_{20} = 12.77$ is larger than the critical value at 1\%.

This analysis only considers time. Figure \ref{fig:prtp} shows that the discount rate used to estimate the social cost of carbon has varied over time. Particularly, the once popular pure rate of time preference of 3.0\% has been largely replaced by 1.5\%. This would increase estimates of the social cost of carbon. Table \ref{tab:means} and Figure \ref{fig:discount} confirm this. Of course, economists using a different discount rate does not mean that the discount rate itself has changed.\cite{Drupp2018}

Table \ref{tab:chi} therefore repeats the analysis for the four pure rates of time preference for which there are observations in every time period: 0\%, 1\%, 2\% and 3\% (see Table \ref{tab:count}). Conditional on the pure rate of time preference, the Equality of Proportions test finds statistically significant differences between the publication periods, except for the lowest discount rate. The Kolmogorov-Smirnov test finds differences at a finer resolution but not for quintiles (see Table \ref{tab:ks}). Note that these tests does not reveal \emph{how} things have changed, only that they have.

Five more analyses are included in the Supplementary Material. These analyses assume normality of error terms, avoiding the asymmetry and thick tail that are a feature of the social cost of carbon but make upward trends harder to detect. The first additional analysis is a weighted linear regression of the social cost of carbon on the pure rate of time preference, which shows a highly significant coefficient, and the year of publication, which is insignificant; either result is independent of the weights used. See Table \ref{tab:regress}. In a second analysis, the pure rate of time preference is dropped. The time trend is still insignificant. In the third analysis, the linear time trend is replaced by a flexible time trend. Again, the effect of publication year is insignificant regardless of weights. See Figure \ref{fig:yearfe}. Fourthly, quantile regression is used. The pure rate of time preference is significant for almost all quantiles and weights; the year of publication for almost none. However, the social cost of carbon appears to increase over time if estimates are weighted for quality and attention is restricted to the central parts of the distribution. See Table \ref{tab:regress}. Fifthly, the pure rate of time preference is dropped again. A trend appears in the 30th percentile, and in the median and 90th percentile if quality weights are used. All together, the upward trend in Figure \ref{fig:means} is partly because analysts have used lower discount rates but also because higher-quality studies have become more pessimistic about climate change.

Earlier claims of an increase in the social cost of carbon\cite{Bergh2014, Nature2017, Hansel2020, Wagner2021b} are confirmed. There is an apparent upward trend because estimates are reported for later years, because of price inflation, and because later analyses tend to use lower discount rates. Correcting for these factors and properly accounting for the asymmetric, heavy-tailed uncertainty, there is indeed a statistically significant time trend in published estimates of the social cost of carbon.

Table \ref{tab:kernelmeans} shows how much estimates of the social cost of carbon have changed. Between earlier periods, estimates went up and down. In the last 10 years, however, there has been a steady increase. For a pure rate of time preference of 3\%, 2\% or 0\%, the estimates more than quadrupled in 2018-2022 compared to 2007-2013. For all estimates, the increase is smaller while estimates decrease for a 1\% PRTP. Overall, however, economists have become more pessimistic about climate change and its impacts.

Research continuously refines our knowledge and updates our estimates, sometimes upwards, sometimes downwards. There are still many things left to research\cite{Burke2016} and things that have been studied but are not yet reflected in our estimates of the social cost of carbon.\cite{NAS2017, Wagner2021a} Researchers will continue to disagree on how best to express the ethical dimensions of the social cost of carbon. Moreover, the social cost of carbon reflects the impact of future climate change and the future will remain uncertain.

The implication of the meta-analysis presented here is that the literature on the social cost of carbon does justify an upward revision of carbon prices and emission reduction targets. Furthermore, the literature, summarized in Table \ref{tab:means}, suggests that, in most of the world, the price of carbon is too low. This is partly because other kinds of climate policies suppress the price of emission permits or reduce the need for a carbon tax. Almost 80\% of greenhouse gas emissions is not priced at all.\cite{WorldBank2021}. Only the EU ETS, now with permit prices around \$250/tC, is in the ballpark of published estimates of the social cost of carbon.\cite{ICAP2021} There is often a gap between the announced emissions targets and the policies supposed to achieve the targets.\cite{UNFCCC2021} Besides raising the social cost of carbon, the \emph{recommended} carbon price, policy makers should focus on raising the \emph{actual} price of carbon.

\begin{table}[hp]
    \centering
    \begin{tabular}{|c|c|c|} \hline
         \textbf{PRTP} & \textbf{empirical} & \textbf{kernel} \\ \hline
         \textbf{3\%} & 43 & 81 \\
         & (54) & (79) \\ \hline
         \textbf{2\%} & 238 & 663 \\
         & (492) & (641) \\ \hline
         \textbf{1\%} & 155 & 420 \\
         & (318) & (502) \\ \hline
         \textbf{0\%} & 407 & 841 \\
         & (539) & (806) \\ \hline
         \textbf{all} & 179 & 509 \\
         & (382) & (560) \\\hline
    \end{tabular}
    \caption{Empirical and kernel average (standard deviation) of estimates of the social cost of carbon (\$/tC) by pure rate of time preference (PRTP).}
    \label{tab:means}
\end{table}

%\begin{table}[hp]
%    \centering
%    \include{chi}
%    \caption{Test for the equality of quintiles between publication periods for the whole sample and for selected pure rates of time preference.}
%    \label{tab:chi}
%\end{table}

\begin{table}[hp]
    \centering
    \begin{tabular}{|l|r|r|r|r|r|}
\hline
&\textbf{Test statistic}&\textbf{p-value}&\textbf{10\%}&\textbf{5\%}&\textbf{1\%}\\\hline
\textbf{All}&11.82&0.92&5.83&6.86&9.51\\\hline
\textbf{PRTP = 0\%}&1.05&1.00&3.23&4.02&5.47\\\hline
\textbf{PRTP = 1\%}&7.88&0.99&3.68&4.85&6.47\\\hline
\textbf{PRTP = 2\%}&15.07&0.77&3.21&4.08&6.15\\\hline
\textbf{PRTP = 3\%}&14.59&0.80&2.12&3.24&5.21\\\hline
\end{tabular}

    \caption{Test for the equality of quintiles between selected publication periods and the whole record, for the whole sample and for selected pure rates of time preference. The third column shows the p-value of the asymptotic $\chi_{20}^2$ test, the next three columns bootstrapped critical values.}
    \label{tab:chi}
\end{table}

\begin{table}[hp]
    \centering
    \begin{tabular}{l r r r r r r} \hline
PRTP & 1982-1995 & 1996-2001 & 2002-2006 & 2007-2013 & 2014-2017 & 2018-2022 \\ \hline
any & 847 &	408 & 277 & 328 &	373 & 683 \\
3\% & 26 & 129 & 36	& 33 &	64 & 146 \\
2\% & 44 & 60	& 74 &	195 &	257 &	1315 \\
1\% & 686 &	118 &	110 &	450 &	264 &	340 \\
0\% & 573 &	1097 &	639 &	446 &	979 &	1925 \\ \hline
\end{tabular}
    \caption{The kernel mean of social cost of carbon (\$/tC) by publication period and pure rate of time preference (PRTP). Estimates are quality weighted and censored.}
    \label{tab:kernelmeans}
\end{table}

\section*{Methods}
\subsection*{Kernel density decomposition}
\label{sc:decomp}
A \textit{kernel density} is defined as
\begin{equation}
\label{eq:kernel}
    f(x) = \frac{1}{nh} \sum_{i=1}^n K \left ( \frac{x-x_i}{h} \right )
\end{equation}
where $x_i$ are a series of observations, $h$ is the bandwidth, and $K$ is the \textit{kernel function}. The kernel function is conventionally assumed to be a (\textit{i}) \textbf{non-negative} (\textit{ii}) \textbf{symmetric} function that (\textit{iii}) \textbf{integrates to one}, with (\textit{iv}) \textbf{zero mean} and (\textit{v}) \textbf{finite variance}.\cite{Takezawa2005} That is, any standardized symmetric probability density function can serve as a kernel function. The Normal density is a common choice.

Conventions are just that. As long as the kernel function is non-negative (the assumption of non-negativity is relaxed for bias reduction.\cite{Jones1997}) and integrates to one, an appropriately weighted sum of kernel functions is non-negative and integrates to one\textemdash such a sum is a probability density function.\cite{Quetelet1846, Quetelet1852, Pearson1894}

The kernel density is thus defined as the sum of probabilities; see Equation (\ref{eq:kernel}). It is a vote-counting procedure\cite{Laplace1814} where the votes are uncertain. This interpretation fits the nature of the data. Estimates of the social cost of carbon are not ``data'' in the conventional sense of the word, nor can integrated assessment models be seen as ``data generating processes''. Besides, I use the population of estimates rather than a sample. There is therefore no Frequentist interpretation of the proposed method. There is no Bayesian interpretation either. While we might take the kernel function to express degrees of belief, a Bayesian procedure would take the first estimates\cite{Nordhaus1982} as prior and later estimates as likelihoods, \emph{multiplying} rather than \emph{adding} probability densities. Because some of the estimates are mutually exclusive with other estimates, a Bayesian interpretation is problematic. In this interpretation, dependence between studies and estimates are not an issue: The repetition of a previous estimate raises confidence in that estimate.

A kernel density can also be seen as a mixture density.\cite{Makov2001, McLachlan2001}. This reinterpretation opens a route to decomposition. We can construct the kernel density of any subset of $x_i$. The weighted sum of the kernel densities of all subsets is a kernel density.

With the right weights and bandwidths, the weighted sum of the kernel densities of subsets of the data is \emph{identical} to the kernel density of the whole data set. To see this, partition the observations into $m$ subsets of length $m_j$ with $\sum_j m_j = n$, as $x_1, ... x_{m_1}, x_{m_1+1}, ..., x_{m_1+m_2}, x_{m_1+m_2+1}, ..., x_n$. Then
\begin{equation}
\label{eq:decomp}
    f(x) = \sum_{j=1}^m \frac{m_j}{n} \frac{1}{m_j h} \sum_{i=\sum_{k=1}^{j-1} m_k +1}^{\sum_{k=1}^{j} m_k} K \left ( \frac{x-x_i}{h} \right ) =: \sum_{j=1}^m \frac{m_j}{n} f_j(x)
\end{equation}
In the middle expression, the inner sum is a sum of kernel functions. The sum is over a subsample of the data. The outer sum is over all subsamples. As the weights and bandwidths are the same, the resulting kernel density is identical to Equation (\ref{eq:kernel}). Moreover, as shown by the right-most expression, each of the components $f_j$ of the composite kernel density $f$ is itself a kernel density. This is the kernel density for a subsample of the data.

Kernel decomposition works with any set of weights that add to one, and with any kernel function or bandwidth for the subsets:
\begin{equation}
\label{eq:mixture}
    f(x) = \sum_{j=1}^m \frac{w_j}{m_j h_j} \sum_{i=\sum_{k=1}^{j-1} m_k +1}^{\sum_{k=1}^{j} m_k} K_j \left ( \frac{x-x_i}{h_j} \right ) =: \sum_{j=1}^m w_j f_j(x;h_j)
\end{equation}
In this case, the composite kernel density is not the same as the kernel density fitted to the complete data set. It is hard to argue for different kernel functions $K_j$ for different subsets of the data, but different subsets of the data would have different spreads and hence bandwidths $h_j$. I do not do this here, instead use the same bandwidth for every subsample.

\subsection*{Inference}
\label{sc:test}
Equation (\ref{eq:mixture}) holds that the kernel density $f(x)$ is composed of $m$ kernel densities $f_j(x)$ with weight $m_j / n$. For each interval $\underline x < x < \bar x$, I test whether the shares of the component densities equal its weight, using the Equality of Proportions test\cite{Pearson1900} but using the bootstrapped distribution rather than the asymptotic $\chi^2$ distribution proposed by Pearson. Suppose, for example, that a component density makes up 17\% of the overall density. Then, the null hypothesis is that the left-tail, central part and right-tail of the component density also make up 17\% of the left-tail, central part and right-tail of the overall density.

Let intervals correspond to $p$ percentiles of the composite distribution. The test statistic is
\begin{equation}
\label{eq:pearson}
    \chi_{(m-1)(p-1)}^2 = \frac{n}{p} \sum_{k=0}^p \sum_{j=1}^m \frac{\left( \int_{P_k}^{P_{k+1}} f_j(x) \mathrm{d} x - \frac{m_j}{n} \right )^2}{\frac{m_j}{n}}
\end{equation}
The test only works if there are two components or more, $m \geq 2$. If not, there would be nothing to compare. The distribution needs to be split in two quantiles or more, $p \geq 2$, because each component density adds up to its weight $m_j / n$ by construction. Again, there would be nothing to compare with fewer than two quantiles.

\printbibliography
\end{refsection}
%\begin{comment}
\begin{refsection}
\newpage \section*{Supplementary materials: Data and methods}
\setcounter{page}{1}
\renewcommand{\thepage}{S\arabic{page}}
\setcounter{table}{0}
\renewcommand{\thetable}{S\arabic{table}}
\setcounter{figure}{0}
\renewcommand{\thefigure}{S\arabic{figure}}
\setcounter{equation}{0}
\renewcommand{\theequation}{S\arabic{equation}}

\subsection*{Papers used in the meta-analysis}
The database draws on earlier meta-analyses of the social cost of carbon,\cite{Tol2005, Tol2009, Tol2010, Tol2011, Tol2013, Tol2018} extended with recent papers that were found using search engines and a review of the publication records of active researchers. 207 papers were used.\cite{Nordhaus1982, Ayres1991, Nordhaus1991EJ, Nordhaus1991AER, Cline1992, Haraden1992, Hohmeyer1992, Nordhaus1992, Penner1992, Haraden1993, Nordhaus1993, Parry1993, Peck1993, Reilly1993, Azar1994, Fankhauser1994, Nordhaus1994, Maddison1995, Schauer1995, Azar1996, Downing1996, Hohmeyer1996, Hope1996, NordhausYang1996, Plambeck1996, Cline1997, NordhausPopp1997, Howarth1998, Eyre1999, Roughgarden1999, Tol1999, NordhausBoyer2000, Tol2000, Kelly2001, Clarkson2002, Mendelsohn2003, Newell2003, Pearce2003, Tol2003, Uzawa2003, Cline2004, Hohmeyer2004, Link2004, Manne2004, Mendelsohn2004, Newell2004, Ceronsky2005, Downing2005, Hope2005a, Hope2005b, Tol2005ede, Guo2006, Hope2006a, Hope2006b, Hope2006c, Stern2006, Wahba2006, Nordhaus2007jel, Nordhaus2007reep, Nordhaus2007sci, Stern2007, Hope2008a, Hope2008b, Nordhaus2008, Anthoff2009ee, Anthoff2009erl, Anthoff2009ejrn, EPA2009, Narita2009, Anthoff2010, IAWGSCC2010, Kemfert2010, Narita2010esp, Narita2010jepm, Newbold2010, Nordhaus2010, Sohngen2010, Tol2010sscc, Anthoff2011, Anthoff2011wp, Ceronsky2011, Hope2011, Marten2011, Nordhaus2011, Pycroft2011, Waldhof2011, AckermanMunitz2012, AckermanStanton2012, Botzen2012, Cai2012, Johnson2012, Kopp2012, Marten2012,  Perrissin2012, Ploeg2012, Rezai2012, Ackerman2013, Anthoff2013, Foley2013, Hope2013, HopeHope2013, IAWGSCC2013, Marten2013, Newbold2013, Nordhaus2013, PloegZeeuw2013, Tol2013el, Weitzman2013, Crost2014, Golosov2014, Heal2014, Howarth2014, Jensen2014, Lemoine2014, Marten2014, Moyer2014, Newbold2014, Nordhaus2014, Ploeg2014, PloegZeeuw2014, Pycroft2014, Waldhoff2014, Cai2015, Dennig2015, Dietz2015, Freeman2015ej, Freeman2015, Hatase2015, Lemoine2015, Lontzek2015, Marten2015, Moore2015, Nordhaus2015, Ploeg2015, Pottier2015, Rezai2015, Shindell2015, Ackerman2016, Bijgaart2016, Cai2016, Freeman2016, Lemoine2016, PloegZeeuw2016, RezaiPloeg2016, Adler2017, Budolfson2017, Dayaratna2017, Golub2017, Hafeez2017, Moore2017, Nordhaus2017, Pindyck2017, PloegRezai2017, RezaiPloeg2017, RezaiPloeg2017ere, RezaiPloeg2017ms, Rose2017, Scovronick2017, Shindell2017, Barrage2018, Ekholm2018, Faulwasser2018, Guivarch2018, Hansel2018, Kotchen2018, Nordhaus2018aej, Nordhaus2018cc, PloegZeeuw2018, Quiggin2018, Ricke2018, Yang2018, Zhen2018, Anthoff2019, Barrage2019, Budolfson2019, Bretschger2019, Cai2019, Daniel2019, Jaakkola2019, Nordhaus2019aer, Nordhaus2019pnas, Pindyck2019, PloegRezai2019, PloegRezai2019eer, Tol2019, Zhen2019, Bastien-Olvera2020, Dayaratna2020, Gschnaller2020, Hansel2020, Howard2020, Kalkuhl2020, Naeini2020, Okullo2020, Scovronick2020, Zhen2020, Bremer2021, Coleman2021, Dietz2021, Drupp2021, Hambel2021EER, Hambel2021JIE, Kikstra2021, Kotlikoff2021, Lupi2021, Taconet2021} The record is close to complete for papers published before 2022.

Most of the papers report results in tabular format. Some only show results in graphs, but most authors emailed the underlying data upon request; if not, the Matlab routine \textsc{grabit}\cite{grabit} was used to digitize the graphs.

The meta-analysis uses the estimate of the social cost of carbon, the year of emission, the year of the nominal dollar, the year of publication, the author, weights, censoring, and the pure rate of time preference. These data, plus some not used here, can be found on \href{https://github.com/rtol/KernelDecomposition/}{GitHub.}

Other factors are known to also affect the social cost of carbon, notably the scenario used, the damage function assumed, and the curvature of the utility function. However, reporting is haphazard and the relationship with the social cost of carbon is complicated. I therefore omit these factors. 

\subsection*{Descriptive statistics}
Table \ref{tab:count} shows the number of estimates per publication period and discount rate, and the number of papers per period. Three papers are counted double because they present comparative results of two different models. There is an upward trend in the number of papers per period, and in the number of estimates per paper.

Table \ref{tab:summstat} shows the mean and standard deviation of the estimates of the social cost of carbon by publication period and year. The range of estimates has grown very wide in recent years.

Figure \ref{fig:period} (top panel) shows the histogram of the published estimates of the social cost of carbon, using quality weights and censoring (see below). Some estimates are negative, a social \emph{benefit} of carbon, but the vast majority is positive. The mode lies between 0 and \$50/tC, but there is a long right tail. The 95th percentile is \$800/tC.

\subsection*{Bandwidth and kernel function}
\label{app:kernel}
\begin{figure}[hp]
    \centering
    \includegraphics[width=\textwidth]{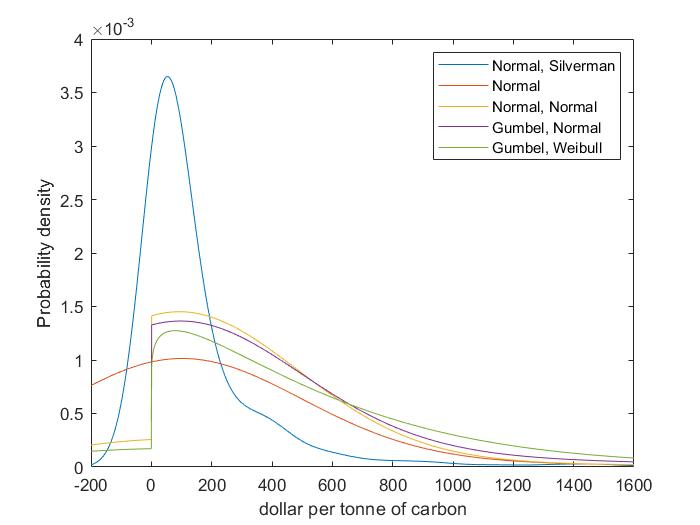}
    \caption{Kernel density of the social cost of carbon for alternative kernel functions and bandwidths.}
    \label{fig:kernel}
\end{figure}

The choice of kernel function and bandwidth is key to any kernel density estimate, as illustrated in Figure \ref{fig:kernel}. If kernel density estimation is interpreted as vote-counting, kernel and bandwidth should be chosen to reflect the nature of the data, shown in Figure \ref{fig:period} (top panel). In this case, the uncertainty about the social cost of carbon is large and right-skewed. Furthermore, the social cost of carbon is, most researchers argue, a cost and not a benefit.

A conventional choice would be to use a Normal kernel function, with a bandwidth according to the Silverman rule\cite{Silverman1986}, that is, 1.06 times the sample standard deviation divided by the fifth root of the number of observations. Figure \ref{fig:kernel} reveals two problems with this approach: The right tail of the resulting kernel density is thin, and a large probability mass is assigned to negative social costs of carbon. See Table \ref{tab:kernstat}. If the bandwidth instead equals the sample standard deviation to reflect the wide uncertainty, the right tail appropriately thickens but the probability of a Pigou subsidy on greenhouse gas emissions increases too.

In order to reflect the skew in the data, I transform the observations. As the social cost of carbon can be both positive and negative, I use the inverse hyperbolic sine. I assume a Normal kernel density with the transformed observation as its mean, mode and median and the sample standard deviation. In the dollar-per-tonne-of-carbon space, this is an arcsinh-normal kernel or Johnson's $S_U$ distribution.\cite{Johnson1949} As shown in Table \ref{tab:kernstat}, this has two effects: The left tail, represented by the probability of a negative social cost of carbon, thins too much and right tail, represented by the chance of a social cost of carbon in excess of the Leviathan tax, thickens too much. The expected social cost of carbon escalates.

With conventional approaches less than satisfactory, unconventional ones should be tried. Many of the published estimates of the social cost of carbon are based on an impact function that excludes benefits of climate change. Climate change can only do damage and additional carbon dioxide can only be bad. Honouring that, I assign a \emph{knotted} Normal kernel function to these observations. As the kernel function is \textbf{asymmetric}, centralization needs to be carefully considered\textemdash is $x_i$ in Equation (\ref{eq:kernel}) the mean, median or mode of $K$? I prefer to use the mode as its central tendency, in line with the interpretation of kernel density estimation as vote-counting (see above). With these assumptions, Figure \ref{fig:kernel} shows that the probability of a negative social cost of carbon falls.

The studies that report the possibility of a negative social cost of carbon nonetheless argue in favour of a positive one. A symmetric Normal kernel does not reflect that. I therefore replace it with a Gumbel kernel:
\begin{equation}
f(x) =  \frac{1}{\beta} \exp \left (-\frac{x-\mu}{\beta} - \exp \left (-\frac{x-\mu}{\beta} \right ) \right )
\end{equation}
The Gumbel distribution is defined on the real line but right-skewed. Again, I use the mode $\mu$ as its central tendency. This thickens the right tail and thins the left tail.

A knotted Normal kernel is not only peculiar near zero but it also has a thin tail. I therefore replace it with a Weibull kernel:
\begin{equation}
f(x) =  \frac{\kappa}{\lambda} \left ( \frac{x}{\lambda} \right )^{\kappa-1} \exp \left ( \left ( \frac{x}{\lambda} \right )^{\kappa} \right )
\end{equation}
The Weibull is defined on the \emph{positive} real line, $f(x)$ is near zero if $x$ is near zero, and right-skewed. I use the mode $\lambda \left (\sfrac{\kappa-1}{\kappa} \right )^{\sfrac{1}{\kappa}}$ as its central tendency. The right tail of the kernel distribution thickens again.

The bottom line of Table \ref{tab:kernstat} shows the Mean Integrated Square Error, the sum of the squared difference between the empirical frequency and modelled probability, weighted by the empirical frequency. The Normal kernel function with the Silverman bandwidth performs best on this criterion, by construction. However, I give preference to the considerations above and use the Weibull-Gumbel kernel distribution as the default. The large probability of a carbon subsidy weighs particularly heavy against the Normal/Silverman kernel density.

The preferred mix of Weibull and Gumbel kernel functions leads to a kernel density that is discontinuous at zero. This is because the 1911 estimates from 42 studies that allow for climate change benefits are qualitative different from the estimates and studies that do not. The final kernel density is therefore a mixture of a density on the real line and a density on the positive real line. The discontinuity at zero follows inevitably.

The Matlab code can be found on \href{https://github.com/rtol/KernelDecomposition/}{GitHub.}

\subsection*{Weights}
\label{app:weight}
The social cost of carbon is an estimate of the willingness to pay to reduce carbon dioxide emissions. Willingness to pay is limited by ability to pay. Papers that report a willingness to pay that exceeds income forgot to impose a key budget constraint. 282 out of 5,791 estimates violate this, from 11 papers (with one paper\cite{Taconet2021} having 232 disqualified estimates). No paper was completely excluded. If the estimated social cost of carbon were levied as a carbon tax, tax revenue would exceed total income. This is not possible: There would be no private income left. These estimates, exceeding \$7,609/tC (the global average carbon intensity in 2010), are excluded.

Another 1,186 estimates (from 37 papers) are so large that, if levied as a carbon tax, the public sector would grow beyond its global average of 15\% in 2010, even if all other taxes would be abolished. Estimates in excess of this Leviathan tax\cite{Tol2012}, \$1,141/tC, are discounted. The discount is linear, varying between 0 for \$1,141/tC and 1 for \$7,609/tC. At most half of the estimates from a single paper\cite{Quiggin2018} were so discounted. Without these discounts, very large estimates of the social cost of carbon dominate the analysis; cf. Table \ref{tab:summstat}.

\begin{figure}[hp]
    \centering
    \includegraphics[width=\textwidth]{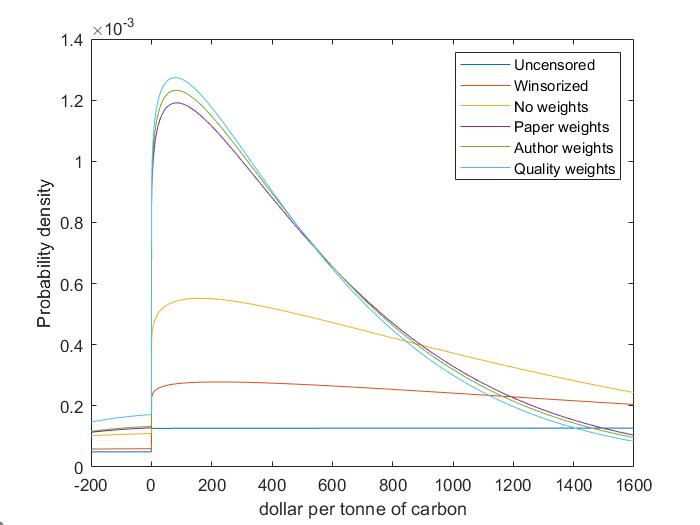}
    \caption {Kernel density of the social cost of carbon for alternative weights.}
    \label{fig:weight}
\end{figure}

The estimates of the social cost of carbon are weighted in four different ways. First, all estimates are treated equally. This is graph ``no weights'' in Figure \ref{fig:weight}. While some papers present a single estimate of the social cost of carbon, other papers show many, up to 1,229 variants\cite{Taconet2021}. This emphasizes studies that ran many sensitivity analyses, which became easier as computers got faster. Therefore, secondly, estimates are weighted such that the total weight \textit{per paper} equals one. This is ``paper weights'' in Figure \ref{fig:weight}. Within each paper, estimates that are favoured by the authors are given higher weight. Favoured estimates are highlighted in the abstract and conclusions, and they are used as the starting point in robustness checks. Estimates that are shown in order to demonstrate that the new model can replicate earlier work are not favoured. I gave double weight to favoured estimates, noting that estimates can be doubly or triply favoured. Some papers report their favoured estimates multiple times, in which case no weights were applied. Authors weights are scaled to add to one. This is ``author weights'' in Figure \ref{fig:weight}.

Different experts have cast different numbers of votes. I count one paper as one vote. One can also argue that it should be one vote per expert, or that papers should be weighted by citations, journal prestige, or author pedigree. Composite kernel densities naturally allow for this, but it is a dangerous route to travel in this case. Older papers are cited more than younger papers; journal rankings are hard enough within disciplines, harder still between disciplines; and prestige, reputation and pedigree often reflect old glory rather than current wits.

Instead, I use a set of weights that reflect the quality of the paper. Over 95\% of the estimates are peer-reviewed; these score 1; the rest score 0. Almost 95\% use an emission scenario; these score 1; papers based on arbitrary emissions score 0 as do papers assuming a steady state. Almost 99\% of estimates estimate the social cost of carbon as a true marginal or a small increment; these score 1; papers with ropy mathematics (e.g. no apparent mathematical model) score 0 as do papers reporting an average rather than a marginal. Over 68\% of estimates assume that vulnerability to climate change is constant; these score 0; papers that recognize that the impacts of climate change vary with development score 1. Only 3\% of estimates is based on new estimates of the total impact of climate change; these score 1; the rest score 0. These scores are added. This is ``quality weights'' in Figure \ref{fig:weight}.

Figure \ref{fig:weight} shows the impact of the weights. The censoring of large estimates, in excess of the Leviathan tax or even in excess of income, has the largest effect. The maximum estimate in the data is \$107,260,751/tC. Not even Jeff Bezos could afford to pay such a carbon tax. Including estimates like these, the kernel bandwidth becomes so large that the kernel density becomes almost uniform.

Winsorizing the data has a similar effect. Replacing estimates in excess of \$7,609/tC by \$7,482/tC, the largest estimate below the threshold, leads to a large kernel bandwidth and an almost uniform kernel density.

The censored data reveal an articulated probability density function. The unweighted estimates have the fattest tail; that is, papers that are more pessimistic about climate change present more estimates of the social cost of carbon. Giving every paper, rather than every estimate, a unit weight thins the right tail. Discounting estimates that their own authors discount further thins the tail. This result is mechanical, as the convention in sensitivity analysis is to show both high and low alternatives to the central assumptions. Quality weights thin the tail further still. More credible studies are less pessimistic.

\subsection*{Inference}
The statistic used to test the null hypothesis that the kernel distribution is equal to its components is given in Equation (\ref{eq:pearson}). This test statistic was proposed by Karl Pearson.\cite{Pearson1900}. It is the normalized sum of squared deviations of the observed numbers from the expected numbers. The distribution of this statistic is asymptotically $\chi^2$ and, for tables greater than $2 \times 2$, relies on a series of approximations. The distinctly non-normal distribution of the social cost of carbon, the weights used, and the kernel distribution further decelerate convergence. The power of the asymptotic test is low as a result.

I therefore bootstrap the test statistics, using a 1,000 random draws (with replacement) from the set of reported social costs of carbon and randomly splitting the sample into 5 (discount rate, author) or 6 (period) groups.

\begin{figure}[hp]
    \centering
    \includegraphics[width=\textwidth]{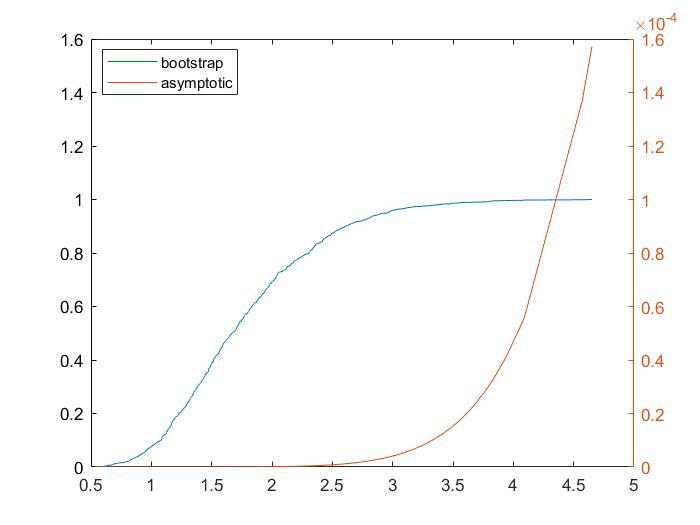}
    \caption{The bootstrapped cumulative distribution function of Pearson's test statistic and its asymptotic p-value.}
    \label{fig:bootstraptest}
\end{figure}

Figure \ref{fig:bootstraptest} contrasts the bootstrapped cumulative density function of Pearson's test statistic and its asymptotic counterpart for the publication periods, the sample split of interest. The asymptotic test is severely underpowered, not rejected the null hypothesis when it should in almost all cases.

\subsection*{Uncertainty about uncertainty}
\begin{figure}[hp]
    \centering
    \includegraphics[width=\textwidth]{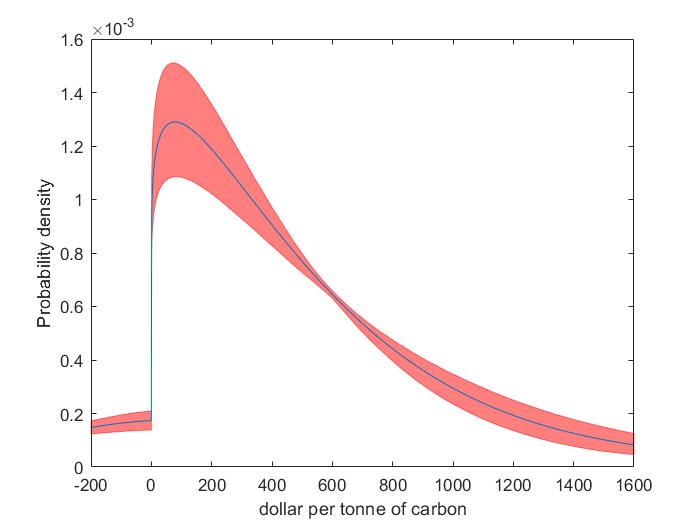}
    \caption{The 95\% confidence interval of the kernel density of the social cost of carbon.}
    \label{fig:bootstrap}
\end{figure}

The kernel density describes the uncertainty about the social cost of carbon. The kernel density is an estimate and as such uncertain. Above and below, I ignore the uncertainty about the uncertainty. Figure \ref{fig:bootstrap} shows the 95\% confidence interval around the Weibull-Gumbel kernel distribution of Figure \ref{fig:kernel}, using quality weights as in Figure \ref{fig:weight}. This confidence interval is based on a bootstrap of 1,000 replications of the published estimates (without reweighing).

The shape of the kernel distribution is well-defined. The key uncertainty is about the weight of the tail relative to the central part of the distribution.

Ignoring the uncertainty about the uncertainty makes it \emph{more} likely to detect patterns. The bootstrap test discussed above therefore includes the meta-uncertainty.

\subsection*{Publication bias}
Publication bias in the literature on the social cost of carbon has been reported \cite{Havranek2015}. Standard tests for publication bias are designed for statistical analyses, particularly test for a reluctance to publish insignificant results. The reported ``publication bias'' thus reflects that few studies report \emph{negative} estimates of the social cost of carbon. Indeed, in the original DICE model \cite{Nordhaus1992}, the social cost of carbon is positive \emph{by construction} (see above). Many later papers follow in Nordhaus' footsteps. The reported \emph{publication} bias is perhaps better interpreted as \emph{confirmation} bias: Researchers adopted Nordhaus' assumption that climate change cannot have positive impacts.

Earlier tests for publication bias\cite{Begg1994} are inapplicable: Estimates of the social costs of carbon are not published as a statistical result. These papers cannot have been selected on a p-value.

A recently proposed test for publication bias\cite{Andrews2019} is more general and can be applied to non-statistical results. The null hypothesis is that earlier and later studies are drawn from the same distribution. If not, later studies are influenced by earlier results\textemdash a sign of publication bias. The test used here is similar in spirit\textemdash does the distribution change over time?\textemdash but the interpretation is about the knowledge base rather than the publication practice. It is not possible to disentangle changes in knowledge from changes in publication practice.

\newpage \section*{Supplementary materials: Results}

\subsection*{To tax or not to tax}
The social cost of carbon can be estimated along an arbitrary emissions scenario or along the optimal emissions scenario. In the latter case, the social cost of carbon equals the carbon tax imposed. That carbon tax is the Pigou tax. The majority of estimates, 4155, are for the Pigou tax. The other 1750 estimates are for arbitrary emissions. As a carbon tax would reduce emissions, the Pigou tax should be lower. It is, but not significantly so. The empirical means are \$192(23)/tC and \$158(19)/tC. The difference is \$36/tC with a standard error of \$30/tC.

\begin{figure}[hp]
    \centering
    \includegraphics[width=\textwidth]{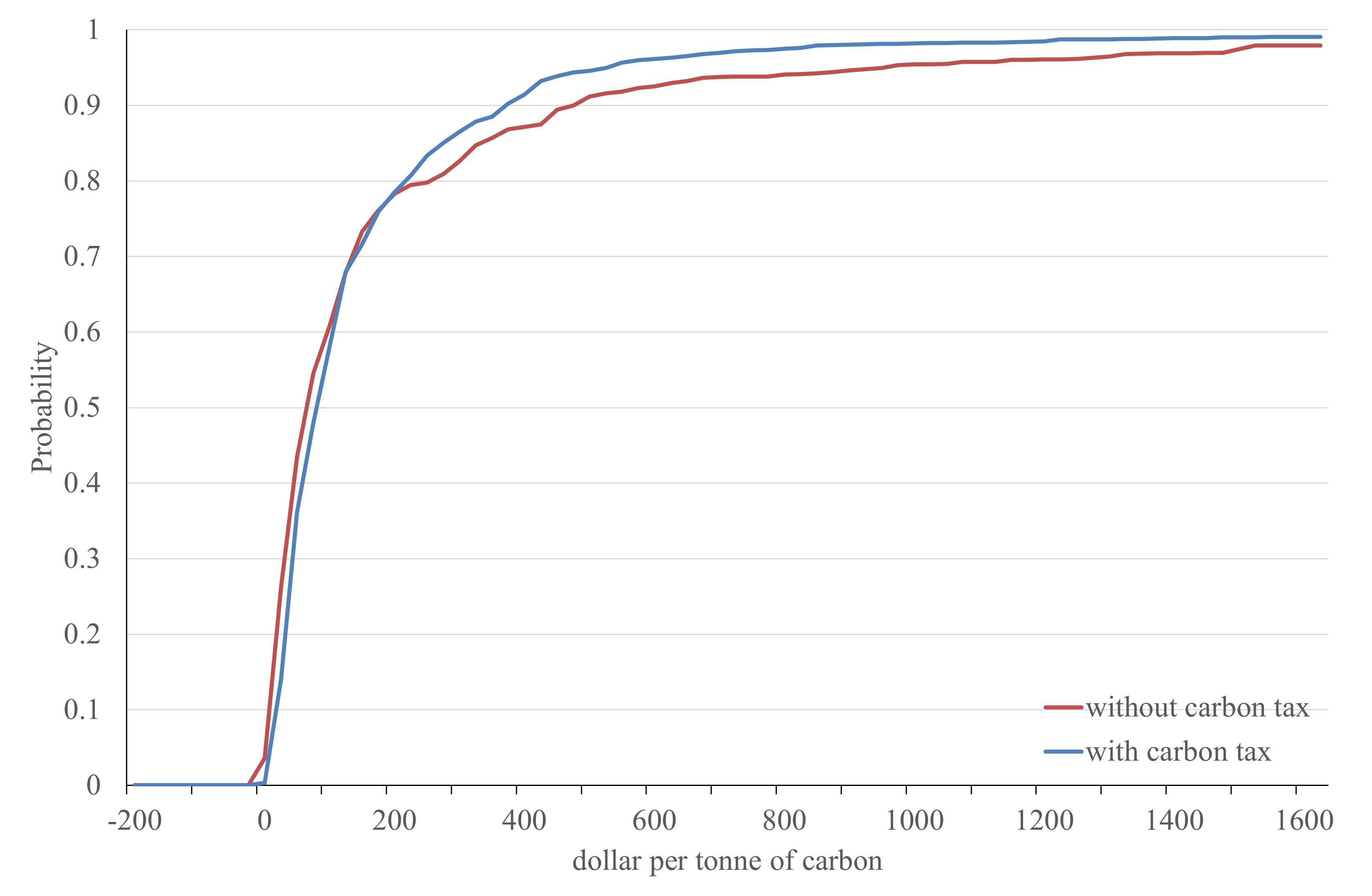}
    \caption{The empirical cumulative distribution function of the social cost of carbon if estimates are (a) along an arbitrary emission scenario and (b) along the optimal emission scenario.}
    \label{fig:pigouscc}
\end{figure}

Figure \ref{fig:pigouscc} shows the empirical cumulative distribution functions. Taxing greenhouse gas emissions takes away some of the heavy right tail of the distribution of the social cost of carbon, but the difference is not that large.

\subsection*{The growth rate of the social cost of carbon}
Different studies report the social cost of carbon for different years of emission. Estimates are standardized to emissions in the year 2010, assuming that the social cost of carbon grows by some 2.2\% per year. That is, estimates for 2000, say, are multiplied by $1.022^{10}$ and estimates for 2020 are divided by the same factor.

\begin{figure}[hp]
    \centering
    \includegraphics[width=\textwidth]{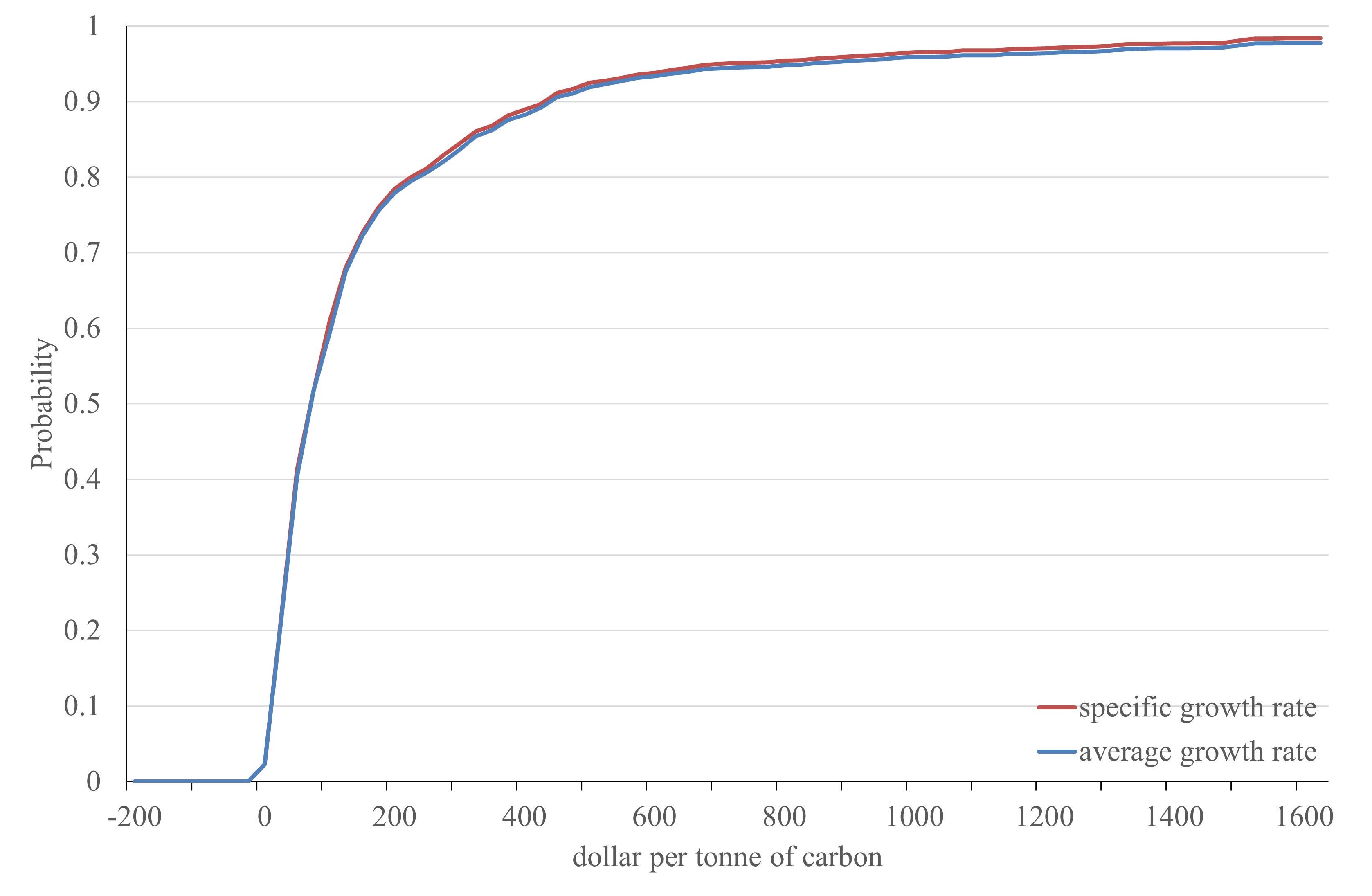}
    \caption{The empirical cumulative distribution function of the social cost of carbon (a) if all estimates are normalized to 2010 with the average growth rather and (b) if estimates are normalized to 2010 with their specific growth rate, if available, or the average growth rate, if not.}
    \label{fig:specificgrowth}
\end{figure}

In the analyses above and below, this correction factor is the same for all studies and for all estimates. This is for comparability and consistency. Alternatively, I use the growth rate associated with that particular estimate, if available, and impute the average growth rate, if not. Figure \ref{fig:specificgrowth} shows the cumulative distribution function for the two alternatives. There is hardly any difference.

\begin{figure}[hp]
    \centering
    \includegraphics[width=\textwidth]{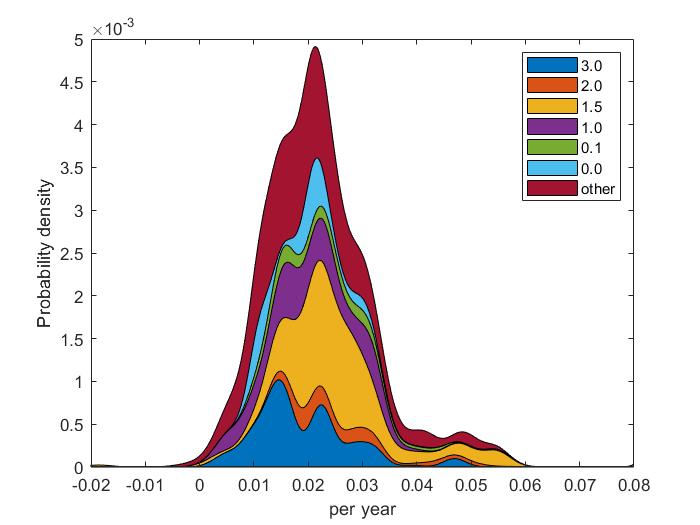}
    \caption{Composite kernel density of the growth rate of the social cost of carbon and its composition by discount rate.}
    \label{fig:growth}
\end{figure}

Figure \ref{fig:growth} shows the composite kernel density of the growth rate of the social cost of carbon, decomposed for the pure rate of time preference; 2.2\% is the mean of this distribution. The density is symmetric for a 3.0\% utility discount rate. However, for discount rates of 1.5\% and 2.0\%, little probability mass is added to the left tail, and a lot to the right tail.

Table \ref{tab:growth} shows the shares by quintile of the kernel density. The same pattern is seen as in the graph, and Pearson's test rejects the null hypothesis that the component densities are equal to the composite one; $\chi^2_{24} = 3.50$; the critical value for 1\% is 2.41. The growth rate of the social cost of carbon differs between discount rates.

However, many estimates of growth rate of the social cost of carbon use a different pure rate of time preference or a different form of discounting altogether. This is the ``other'' in Figure \ref{fig:growth}. The same is true for the estimates of the social cost of carbon itself; see Figure \ref{fig:discount}. A discount-rate-specific growth rate would introduce further assumptions to impute growth rates. I avoid that complexity and assume the same growth rate for all estimates.

\begin{figure}[hp]
    \centering
    \includegraphics[width=\textwidth]{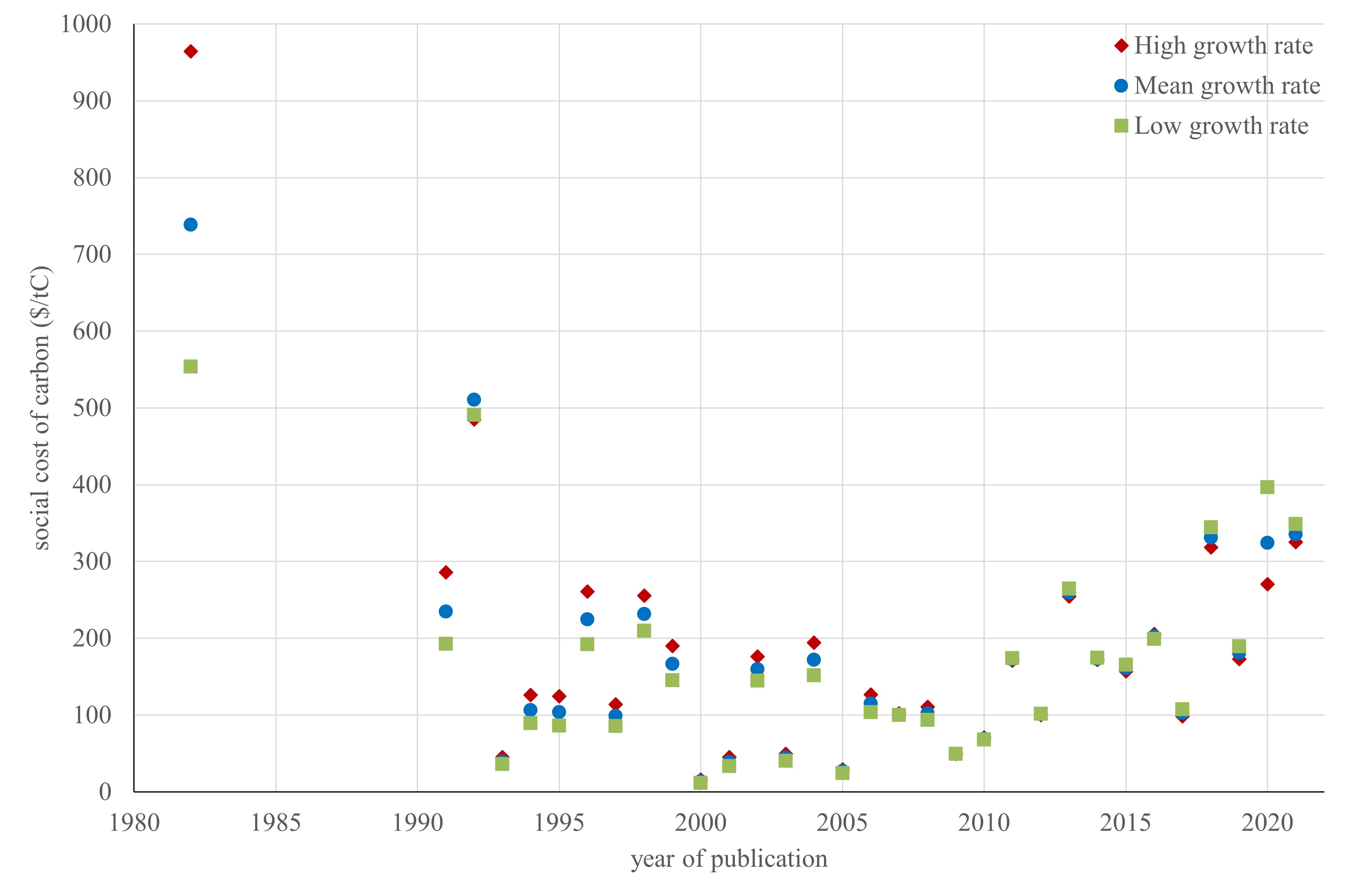}
    \caption{Average social cost of carbon by publication year, corrected for inflation and year of emission, for alternative growth rates of the social cost of carbon: Blue dots use the average of published growth rates, red triangles (green squares) the average plus (minus) the standard deviation. Estimates are quality weighted and censored.}
    \label{fig:growthrate}
\end{figure}

The core analysis assumes that the social cost of carbon grows by 2.2\% per year, the average of the published estimates of the growth rate. Figure \ref{fig:growthrate} shows the quality-weighted and censored means per publication year for two alternative assumptions: The average plus or minus the standard deviation of the published estimates. If the growth rate of the social cost of carbon is 3.2\% per year, then older estimates, normalized to 2010 as the year of emission, are higher and younger estimates are lower. The reverse happens when the growth rate is assumed to be 1.2\% per year. That is, if a faster (slower) growth rate is assumed, an upward trend will be harder (easier) to detect.

This result is paradoxical. The social cost of carbon is a measure of the current seriousness of climate change. The growth rate of the social cost of carbon is a measure of the future seriousness of climate change. The faster the growth of the social cost of carbon, the more serious climate change, but the harder it is to detect an upward trend in the social cost of carbon.

\subsection*{Time and discount}
\begin{figure}[hp]
    \centering
    \includegraphics[width=\textwidth]{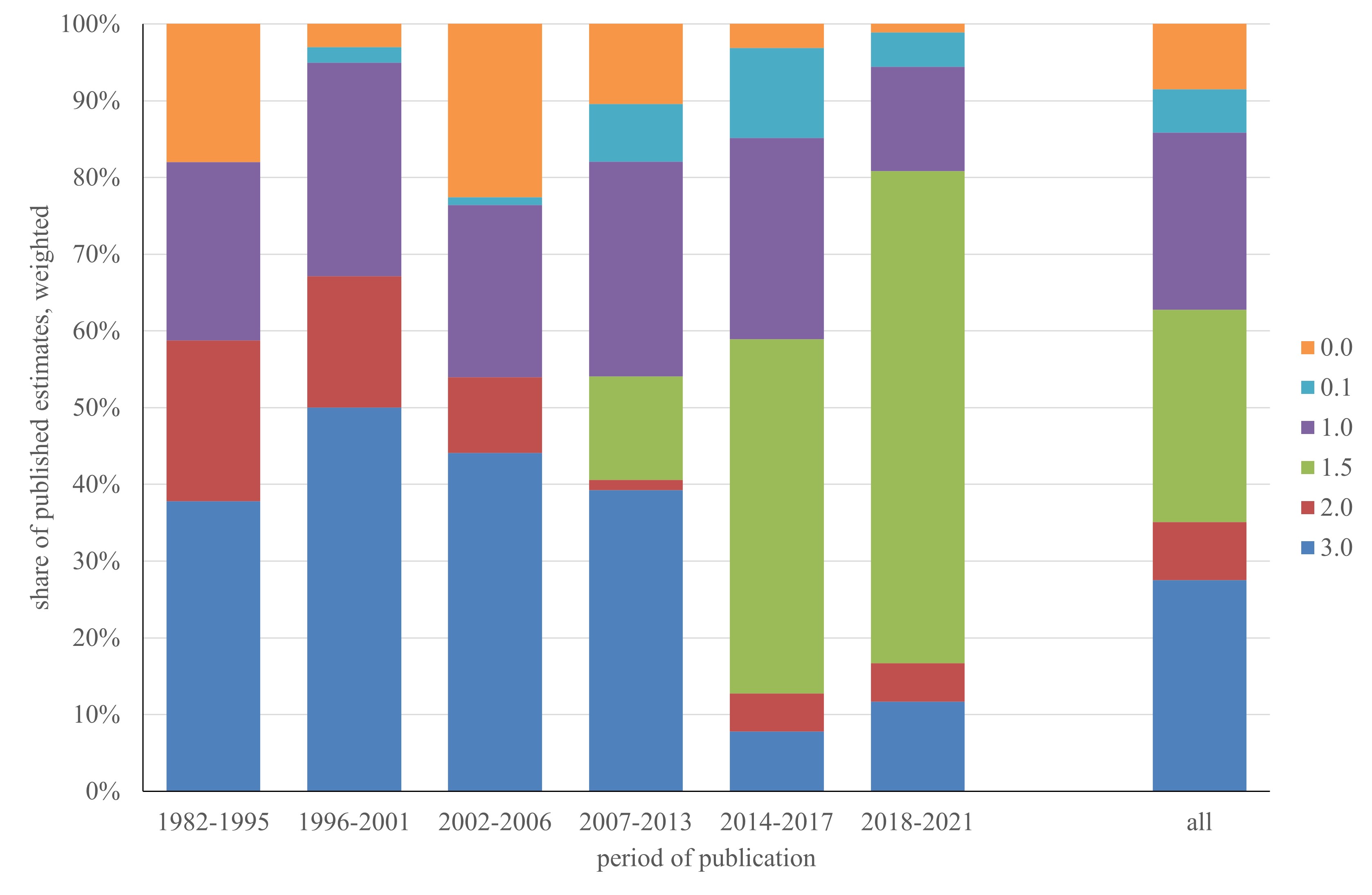}
    \caption{The pure rate of time preference used to estimate the social cost of carbon by publication period. Estimates are weighted such that every published paper counts equally.}
    \label{fig:prtp}
\end{figure}

Table \ref{tab:period} and Figure \ref{fig:period} (bottom panel) show that the distribution of the social cost of carbon has significantly changed over time. Figure \ref{fig:prtp}, however, reveals that the pure rate of time preference used to estimate the social cost of carbon has changed over time. Notably, a 1.5\% pure rate of time preference was first used in 2011 and became the dominant choice in later years. It could be that the increase in the social cost of carbon is because analysts prefer to use a lower discount rate.

\begin{figure}[hp]
    \centering
    \includegraphics[width=\textwidth]{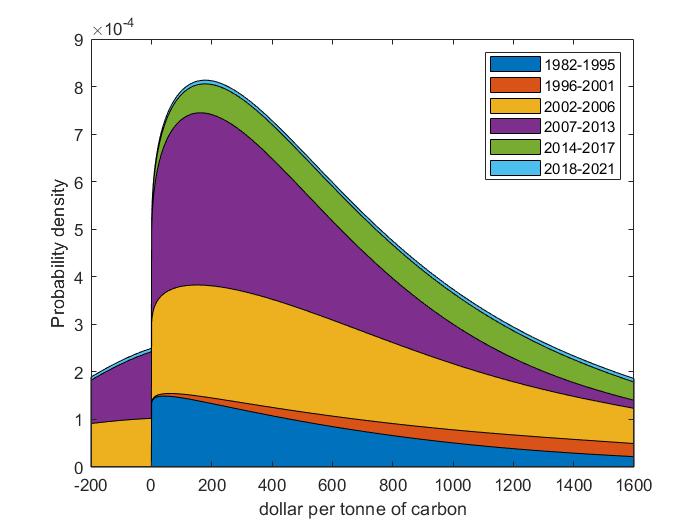}
    \caption{Composite kernel density of the social cost of carbon and its composition by publication period, for a pure rate of time preference of 0\%.}
    \label{fig:period0}
\end{figure}

\begin{figure}[hp]
    \centering
    \includegraphics[width=\textwidth]{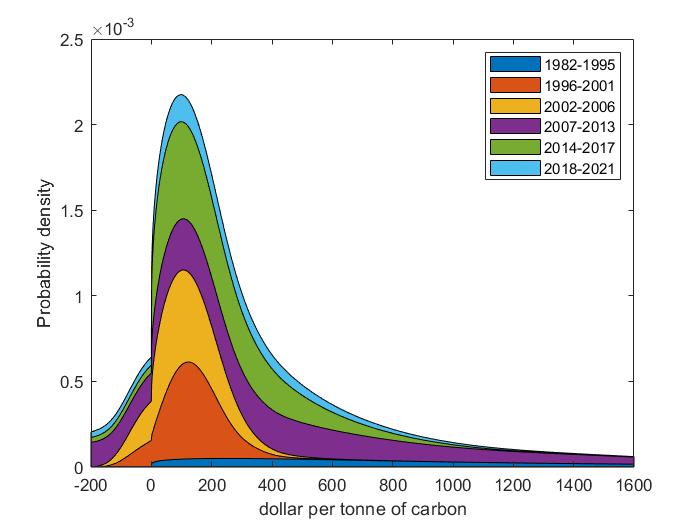}
    \caption{Composite kernel density of the social cost of carbon and its composition by publication period, for a pure rate of time preference of 1\%.}
    \label{fig:period1}
\end{figure}

\begin{figure}[hp]
    \centering
    \includegraphics[width=\textwidth]{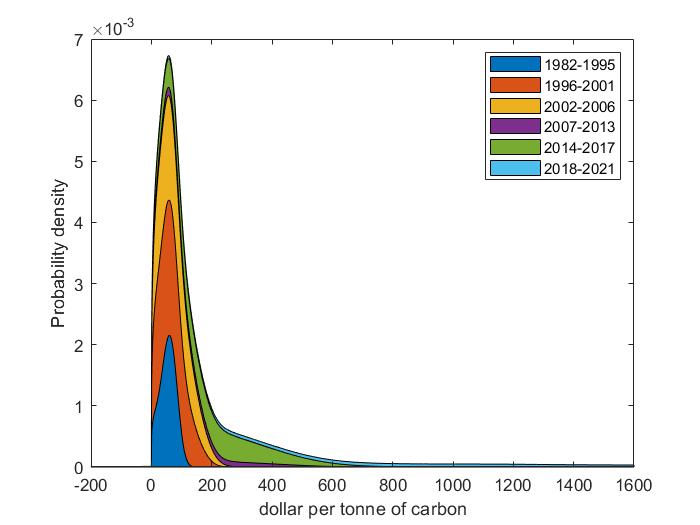}
    \caption{Composite kernel density of the social cost of carbon and its composition by publication period, for a pure rate of time preference of 2\%.}
    \label{fig:period2}
\end{figure}

\begin{figure}[hp]
    \centering
    \includegraphics[width=\textwidth]{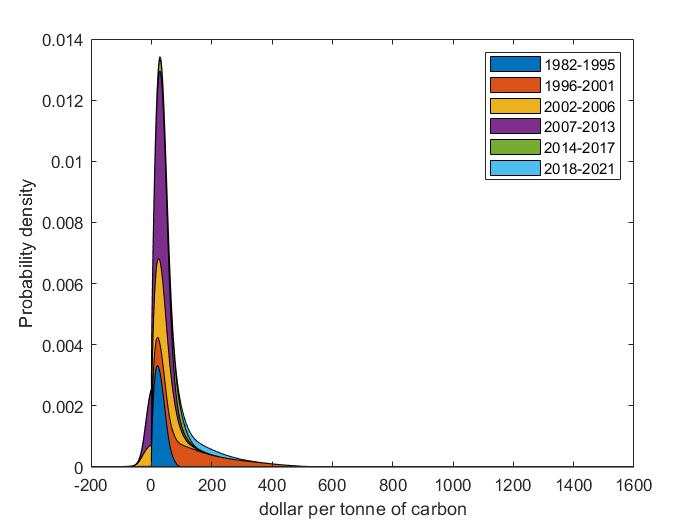}
    \caption{Composite kernel density of the social cost of carbon and its composition by publication period, for a pure rate of time preference of 3\%.}
    \label{fig:period3}
\end{figure}

I therefore redo the decomposition per period for four alternative pure rates of time preference, 0\%, 1\%, 2\%, and 3\%, which have been used throughout the period. Tables \ref{tab:period0}, \ref{tab:period1}, \ref{tab:period2} and \ref{tab:period3} and Figures \ref{fig:period0}, \ref{fig:period1}, \ref{fig:period2} and \ref{fig:period3} show the detailed results, Table \ref{tab:chi} the summary. The null hypothesis that the six periods show the same probability density function for the social cost of carbon cannot be rejected for a 0\% pure rate of time preference\textemdash the uncertainty is so large that any signal is swamped\textemdash but for larger discount rates, differences between periods are statistically significant.

\subsection*{Alternative non-parametric tests}
Pearson's equality of \emph{proportions} test is designed to compare the distributions of subsamples to the distribution of the whole sample. Alternatively, the subsample distributions can be scaled up to sum to one, and tests for the equality of \emph{distributions} can be applied. However, for most of these tests, critical values are tabulated for specific null hypotheses only. These tests can be use to check whether the subsample distribution is, say, Normal but not whether it conforms to the whole-sample kernel distribution. The Kolmogorov-Smirnov test is the exception: Its test statistic converges to a known distribution, independent of the null hypothesis.\cite{Kolmogorov1933} I did not use the standard tabulations\cite{Smirnov1948}, instead computed the p-values.\cite{kolmogorov} Pearson's Equality of Proportions test considers the difference between entire distributions. The Kolmogorov-Smirnov test, on the other hand, consider the maximum deviation between distributions.

Table \ref{tab:ks} shows the results. The null hypothesis that the quintiles of the subperiod distributions are equal to the quintiles of the distribution of the whole period, cannot be rejected. The equality of proportions test was applied to quintiles too. The two statistical tests disagree (although the asymptotic Kolmogorov-Smirnov test agrees with the asymptotic Pearson test).

The null hypothesis cannot be rejected for deciles and ventiles either. However, the null hypothesis is rejected for the third period if quinquagintiles are considered. For centiles, the null hypothesis is rejected for the first, second and fourth periods, but not for the other periods. That is, the subperiod distributions are indistinguishable at a crude resolution but differences appear at a finer scale.

Tables \ref{tab:ks0} to \ref{tab:ks3} repeat the analysis, splitting the sample by discount rate and time period. Rejections of the null hypothesis are common, also for deciles and ventiles, and more so for lower discount rates. However, Table \ref{tab:count} reveals that cell-sizes can be small, down to five estimates. With so few observations, confidence in the estimated kernel densities is low. Nonetheless, the more discerning Kolmogorov-Smirnov test points to changes over time that the asymptotic Pearson test cannot detect. The results of the Kolmogorov-Smirnov test are in line with the bootstrap Pearson test.

\subsection*{Parametric tests}
Table \ref{tab:regress} shows the results for the weighted linear regression, based on the conventional assumptions of linearity and normality, of the social cost of carbon on the pure rate of time preference and the year of publication, using paper, author and quality weights. The time trend is not statistically significant from zero.

\begin{figure}[hp!]
    \centering
    \includegraphics[width=\textwidth]{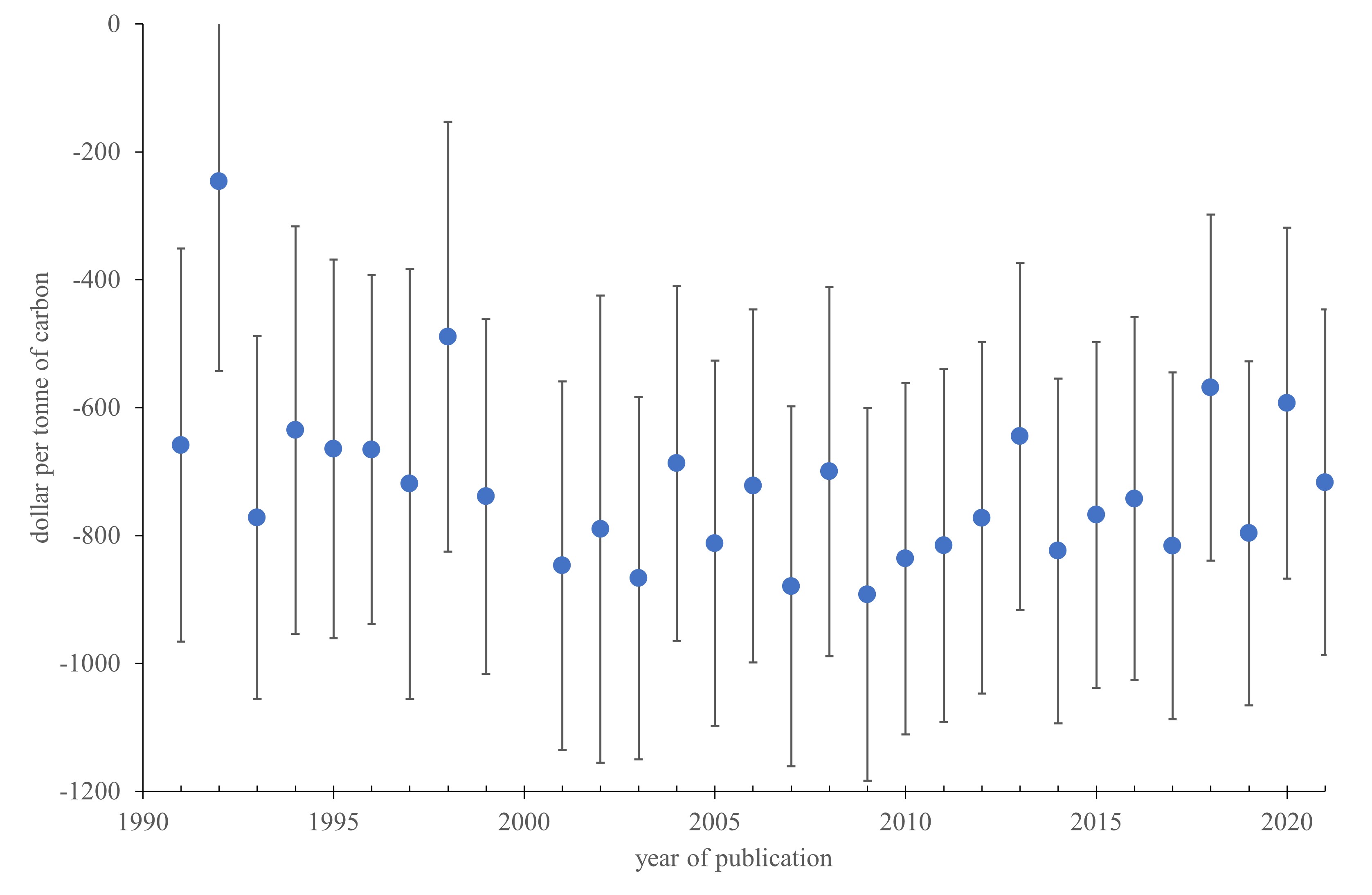}
    \caption{Year fixed-effects from a regression of the social cost of carbon on the pure rate of time preference, using quality weights. Base year is 1982; error bars denote the 67\% confidence interval.}
    \label{fig:yearfe}
\end{figure}

I repeat the analysis using year fixed effects rather than a linear time trend. Figure \ref{fig:yearfe} shows the estimated time dummies, measuring the deviation from 1982. The dummies do not show a trend.

Table \ref{tab:regress} also shows the results of quantile regressions, for quintiles as above. The time trend is insignificant for paper and author weights. Using quality weights, the time trend is positive and significant at the 5\% level for the three lower quintiles. That is, lower estimates of the social cost of carbon are gradually disappearing from the higher-quality literature.

\subsection*{Discount rate}
The social cost of carbon is the net prevent value of the additional future damages done by emitting slightly more carbon dioxide. The assumed discount rate is obviously important in its calculation.

\begin{figure}[hp]
    \centering
    \includegraphics[width=\textwidth]{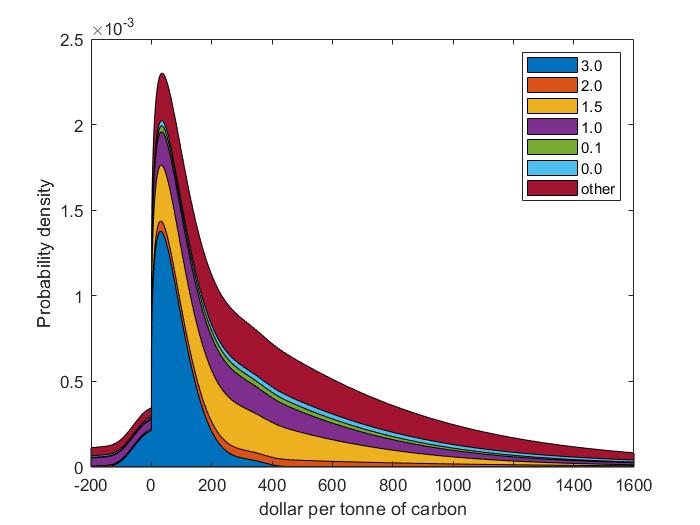}
    \caption{Composite kernel density of the social cost of carbon and its composition by the pure rate of time preference.}
    \label{fig:discount}
\end{figure}

Figure \ref{fig:discount} decomposes the kernel density of the social cost of carbon into its components by pure rate of time preference used. ``Other'' refers to a range of numbers and methods, including constant consumption rates, various forms of declining discount rates, and Epstein-Zin preferences. As one would expect, the lower discount rates contribute more to the right tail of the distribution.

Table \ref{tab:discount} shows the contributions of estimates of the social cost of carbon using a particular pure rate of time preference to the overall kernel density (denoted ``null'') as well as to the five quintiles of that density (denoted Q1-5). The null hypothesis that all shares are equal is firmly rejected; $\chi^2_{24} = 50.69$; the 1\% critical value is 4.79.

This result justifies splitting the sample by discount rate.

\subsection*{Author}
I also test whether different researchers reach different conclusions, one test of the impact of subjective judgements on estimates of the social cost of carbon.

The decomposition of the kernel density by author opens another interpretation of kernel densities: Vote-counting.\cite{Laplace1814} Different experts have published different estimates of the social cost of carbon. These can be seen as votes for a particular Pigou tax. But as the experts are uncertain, they have voted for a central value and a spread. The kernel function is a vote, the kernel density adds those votes. Note the difference with Bayesian updating, which multiplies rather than adds probabilities.

I split the sample into estimates by those who have published ten papers or more (i.e., Christopher W. Hope, William D. Nordhaus, Frederick van der Ploeg, Richard S.J. Tol) and others.

\begin{figure}[hp]
    \centering
    \includegraphics[width=\textwidth]{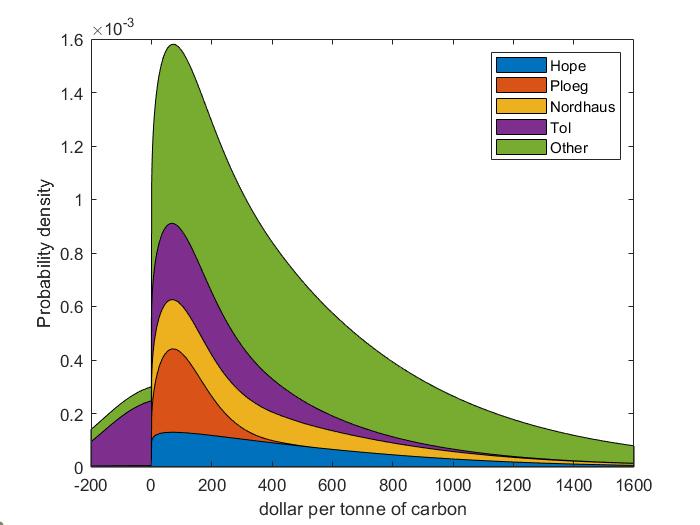}
    \caption{Composite kernel density of the social cost of carbon and its composition by author.}
    \label{fig:author}
\end{figure}

Figure \ref{fig:author} decomposes the kernel density by author. Of the named authors, estimates by van der Ploeg are the narrowest, Tol contributes most to the left tail, and Hope to the right tail.

Table \ref{tab:author} shows the contributions of estimates of the social cost of carbon published by a particular author to the overall kernel density and its quintiles. There are patterns in figure and table, and the null hypothesis that the quintile shares are indistinguishable from the overall shares can be rejected; $\chi^2_{16} = 20.42$; the 1\% critical value is 3.67.

This result notwithstanding, I do not split the sample by period, discount rate, and author because cell sizes would be too small.

%tables
\begin{table}[hp]
    \centering
    \footnotesize\begin{tabular}{|l|c|c|c|c|c|c|c|c}
\hline
&\textbf{1982-1995}&\textbf{1996-2001}&\textbf{2002-2006}&\textbf{2007-2013}&\textbf{2014-2017}&\textbf{2018-2021}&\textbf{total}\\\hline
\textbf{3.0\%}&33 & 27 & 39 &	108 & 24 & 106 & 337\\\hline
\textbf{2.0\%}&5 &	7	& 14 &	7 &	139 &	313 &	485\\\hline
\textbf{1.5\%}&0 &	0	& 0	& 38 & 216 & 1705 & 1959\\\hline
\textbf{1.0\%}&8 & 16 &	28 & 100 & 190 & 448 &	790\\\hline
\textbf{0.1\%}&0 &	4 &	1 &	21 & 69 &	143 & 	238\\\hline
\textbf{0.0\%}&68 &	5 &	33 &	43 &	124 &	262 &	535\\\hline
\textbf{other}&12 &	25 & 26 & 317 &	287 & 894 & 1561\\\hline
\textbf{\# estimates}& 126 &	84 & 141 & 634 & 1049 & 3871 &	5905\\\hline
\textbf{\# papers} &  19 &	18 &	23 &	51 &	52 &	44 &	207\\\hline
\end{tabular}
    \caption{Number of papers on the social cost of carbon by publication period and the number of estimates by period and pure rate of time preference.}
    \label{tab:count}
\end{table}

\begin{table}[hp]
    \centering
    \footnotesize\begin{tabular}{|l|c|c|c|c|c|c|c|c}
\hline
&\textbf{1982-1995}&\textbf{1996-2001}&\textbf{2002-2006}&\textbf{2007-2013}&\textbf{2014-2017}&\textbf{2018-2021}&\textbf{all}\\\hline
\textbf{3.0\%}&25 & 73 & 27 &	25 &	53	& 126 &	45 \\
& (15) & (98) &	(26) & (16) & (27) & (50) &	(56) \\ \hline
\textbf{2.0\%}&46 & 64 & 70 & 136 & 170 & 953 & 221\\
 & (23) &	(32) &	(48) &	(125) &	(137) &	(1185) &	(559)\\\hline
\textbf{1.5\%}& &  & & 196 & 1655 & 11033 & 6380 \\
& & & & (533) & (12318) & (393133)	& (283800) \\\hline
\textbf{1.0\%}& 562 & 107 & 108 & 220 & 203 & 7669 & 1211\\
& (548) & (67) & (117) & (364)	& (977) &	(67058) & (24545) \\\hline
\textbf{0.1\%}& & 671 & 634 & 315 & 447 & 334 & 	400\\
& & (395) & & (418) & (1326) & (7673) & (955)\\\hline
\textbf{0.0\%}&496 & 1054 & 513 & 388 & 1603 & 1369 &	606\\
& (907) & (534) & (642) & (1077) & (3083) & (2633) & (1325) \\\hline
\textbf{other}&1226 & 538 & 131 & 146 & 175 & 22916 & 5053\\
& (2097) & (763) & (123) & (183) & (325) & (1225350) & (562339)\\\hline
\textbf{all} &  481 & 252 & 160 & 157 & 629 & 12849 & 3106\\
& (1212) & (504) & (345) & (387) & (6554) & (736271) & (345471) \\\hline
\end{tabular}
    \caption{Average (standard deviation) of estimates of the social cost of carbon (\$/tC) by publication period and pure rate of time preference. Estimates are uncensored and unweighted.}
    \label{tab:summstat}
\end{table}

\begin{table}[hp]
    \centering
    \scriptsize
\begin{tabular}{|l|c|c|c|c|c|c|c|c|}
\hline
&\thead{\textbf{Normal} \\ \textbf{Silverman}}&\textbf{Normal}&\thead{\textbf{Johnson SU} \\ \textbf{Silverman}}&\textbf{Johnson SU}&\thead{\textbf{Normal} \\ \textbf{Normal}} & \thead{\textbf{Gumbel} \\ \textbf{Normal}}& \thead{\textbf{Gumbel} \\ \textbf{ Weibull}}&\textbf{Observed}\\\hline
\textbf{Average}&179&225&1083&2453&361&4376&509&179\\\hline
\textbf{$P(SCC<0$)}&0.1882&0.3373&0.0005&0.0007&0.0911&0.0660&0.0660&0.0151\\\hline
\textbf{$P(SCC>1186)$}&0.0234&0.0299&0.2845&0.6176&0.0310&0.0611&0.0972&0.0231\\\hline
\textbf{MISE}&0.0396&0.0607&0.0588&0.0674&0.0563&0.0572&0.0586&0.0000\\\hline
\end{tabular}
    \caption{Key characteristics of alternative kernel densities and the data: Mean social cost of carbon (\$/tC), probability of a social benefit of carbon, probability that the social cost of carbon exceeds the Leviathan tax, and Mean Integrated Squared Error (MISE).}
    \label{tab:kernstat}
\end{table}

\begin{table}[hp]
    \centering
    \begin{tabular}{|l|c|c|c|c|c|c|c|}
\hline
&\textbf{3.0}&\textbf{2.0}&\textbf{1.5}&\textbf{1.0}&\textbf{0.1}&\textbf{0.0}&\textbf{other}\\\hline
\textbf{Q1}&0.0464&0.0014&0.0181&0.0293&0.0068&0.0202&0.0751\\\hline
\textbf{Q2}&0.0407&0.0083&0.0383&0.0299&0.0096&0.0070&0.0644\\\hline
\textbf{Q3}&0.0252&0.0107&0.0593&0.0208&0.0059&0.0239&0.0566\\\hline
\textbf{Q4}&0.0234&0.0098&0.0771&0.0208&0.0090&0.0089&0.0507\\\hline
\textbf{Q5}&0.0204&0.0154&0.0708&0.0270&0.0118&0.0066&0.0505\\\hline
\textbf{Null}&0.0312&0.0091&0.0527&0.0256&0.0086&0.0133&0.0595\\\hline
\end{tabular}
    \caption{Observed and hypothesized contribution to the kernel density of the growth rate of the social cost of carbon by quintile and pure rate of time preference.}
    \label{tab:growth}
\end{table}

\begin{table}[hp]
    \centering
    \begin{tabular}{|l|c|c|c|c|c|c|}
\hline
&\textbf{1982-1995}&\textbf{1996-2001}&\textbf{2002-2006}&\textbf{2007-2013}&\textbf{2014-2017}&\textbf{2018-2021}\\\hline
\textbf{Q1}&0.0078&0.0199&0.0328&0.0814&0.0565&0.0284\\\hline
\textbf{Q2}&0.0116&0.0237&0.0270&0.0638&0.0703&0.0297\\\hline
\textbf{Q3}&0.0125&0.0224&0.0225&0.0580&0.0606&0.0324\\\hline
\textbf{Q4}&0.0150&0.0203&0.0170&0.0469&0.0467&0.0367\\\hline
\textbf{Q5}&0.0303&0.0176&0.0092&0.0247&0.0236&0.0507\\\hline
\textbf{Null}&0.0154&0.0208&0.0217&0.0550&0.0515&0.0356\\\hline
\end{tabular}
    \caption{Observed and hypothesized contribution to the kernel density of the social cost of carbon by quintile and publication period.}
    \label{tab:period}
\end{table}

\begin{table}[hp]
    \centering
    \begin{tabular}{|l|c|c|c|c|c|c|}
\hline
&\textbf{1982-1995}&\textbf{1996-2001}&\textbf{2002-2006}&\textbf{2007-2013}&\textbf{2014-2017}&\textbf{2018-2021}\\\hline
\textbf{Q1}&0.0268&0.0014&0.0835&0.1003&0.0092&0.0047\\\hline
\textbf{Q2}&0.0360&0.0050&0.0720&0.0987&0.0217&0.0025\\\hline
\textbf{Q3}&0.0291&0.0089&0.0711&0.0668&0.0272&0.0032\\\hline
\textbf{Q4}&0.0232&0.0157&0.0660&0.0334&0.0326&0.0045\\\hline
\textbf{Q5}&0.0147&0.0237&0.0644&0.0108&0.0267&0.0162\\\hline
\textbf{Null}&0.0260&0.0109&0.0714&0.0620&0.0235&0.0062\\\hline
\end{tabular}
    \caption{Observed and hypothesized contribution to the kernel density of the social cost of carbon by quintile and publication period, for a pure rate of time preference of 0\%.}
    \label{tab:period0}
\end{table}

\begin{table}[hp]
    \centering
    \begin{tabular}{|l|c|c|c|c|c|c|}
\hline
&\textbf{1982-1995}&\textbf{1996-2001}&\textbf{2002-2006}&\textbf{2007-2013}&\textbf{2014-2017}&\textbf{2018-2021}\\\hline
\textbf{Q1}&0.0023&0.0311&0.0507&0.0812&0.0455&0.0225\\\hline
\textbf{Q2}&0.0076&0.0791&0.0754&0.0470&0.0836&0.0237\\\hline
\textbf{Q3}&0.0096&0.0229&0.0237&0.0496&0.0629&0.0200\\\hline
\textbf{Q4}&0.0134&0.0013&0.0021&0.0552&0.0385&0.0155\\\hline
\textbf{Q5}&0.0327&0.0000&0.0002&0.0815&0.0111&0.0101\\\hline
\textbf{Null}&0.0131&0.0269&0.0304&0.0629&0.0483&0.0184\\\hline
\end{tabular}
    \caption{Observed and hypothesized contribution to the kernel density of the social cost of carbon by quintile and publication period, for a pure rate of time preference of 1\%.}
    \label{tab:period1}
\end{table}

\begin{table}[hp]
    \centering
    \begin{tabular}{|l|c|c|c|c|c|c|}
\hline
&\textbf{1982-1995}&\textbf{1996-2001}&\textbf{2002-2006}&\textbf{2007-2013}&\textbf{2014-2017}&\textbf{2018-2021}\\\hline
\textbf{Q1}&0.1507&0.2579&0.2119&0.0205&0.0769&0.0114\\\hline
\textbf{Q2}&0.0000&0.0095&0.0177&0.0167&0.0837&0.0102\\\hline
\textbf{Q3}&0.0000&0.0000&0.0000&0.0080&0.0399&0.0140\\\hline
\textbf{Q4}&0.0000&0.0000&0.0000&0.0004&0.0042&0.0214\\\hline
\textbf{Q5}&0.0000&0.0000&0.0000&0.0000&0.0009&0.0442\\\hline
\textbf{Null}&0.0301&0.0535&0.0459&0.0091&0.0411&0.0202\\\hline
\end{tabular}
    \caption{Observed and hypothesized contribution to the kernel density of the social cost of carbon by quintile and publication period, for a pure rate of time preference of 2\%.}
    \label{tab:period2}
\end{table}

\begin{table}[hp]
    \centering
    \begin{tabular}{|l|c|c|c|c|c|c|}
\hline
&\textbf{1982-1995}&\textbf{1996-2001}&\textbf{2002-2006}&\textbf{2007-2013}&\textbf{2014-2017}&\textbf{2018-2021}\\\hline
\textbf{Q1}&0.0535&0.0199&0.0597&0.1145&0.0057&0.0031\\\hline
\textbf{Q2}&0.0736&0.0268&0.0749&0.1648&0.0129&0.0032\\\hline
\textbf{Q3}&0.0187&0.0280&0.0505&0.0613&0.0206&0.0057\\\hline
\textbf{Q4}&0.0006&0.0348&0.0173&0.0088&0.0156&0.0112\\\hline
\textbf{Q5}&0.0000&0.0842&0.0027&0.0010&0.0010&0.0255\\\hline
\textbf{Null}&0.0293&0.0387&0.0410&0.0701&0.0112&0.0097\\\hline
\end{tabular}
    \caption{Observed and hypothesized contribution to the kernel density of the social cost of carbon by quintile and publication period, for a pure rate of time preference of 3\%.}
    \label{tab:period3}
\end{table}

\begin{table}[hp]
    \centering
    \begin{tabular}{|l|c|c|c|c|c|}
\hline
&\textbf{5}&\textbf{10}&\textbf{20}&\textbf{50}&\textbf{100}\\\hline
\textbf{1982-1995}&0.9991&0.9321&0.5992&0.1039&0.0055\\\hline
\textbf{1996-2001}&0.9903&0.8160&0.3937&0.0350&0.0006\\\hline
\textbf{2002-2006}&1.0000&1.0000&0.9967&0.8032&0.3779\\\hline
\textbf{2007-2013}&0.9999&0.9807&0.7726&0.2218&0.0244\\\hline
\textbf{2014-2017}&1.0000&1.0000&0.9999&0.9488&0.6398\\\hline
\textbf{2018-2021}&1.0000&1.0000&1.0000&0.9808&0.7708\\\hline
\end{tabular}
    \caption{p-values of Kolmogorov-Smirnov test statistic for the equality of the distributions of the period subsample and the whole sample (rows) for different discretizations of the distribution (columns).}
    \label{tab:ks}
\end{table}

\begin{table}[hp]
    \centering
    \begin{tabular}{|l|c|c|c|c|c|}
\hline
&\textbf{5}&\textbf{10}&\textbf{20}&\textbf{50}&\textbf{100}\\\hline
\textbf{1982-1995}&1.0000&0.9780&0.8312&0.2106&0.0164\\\hline
\textbf{1996-2001}&0.1191&0.0064&0.0000&0.0000&0.0000\\\hline
\textbf{2002-2006}&0.9867&0.7959&0.3724&0.0301&0.0005\\\hline
\textbf{2007-2013}&1.0000&0.9998&0.9697&0.5846&0.1781\\\hline
\textbf{2014-2017}&1.0000&0.9780&0.7602&0.1907&0.0152\\\hline
\textbf{2018-2021}&1.0000&1.0000&0.9924&0.7203&0.2647\\\hline
\end{tabular}
    \caption{p-values of Kolmogorov-Smirnov test statistic for the equality of the distributions of the social cost of carbon for the period subsample and the whole sample (rows) for different discretizations of the distribution (columns), for a 0\% pure rate of time preference.}
    \label{tab:ks0}
\end{table}

\begin{table}[hp]
    \centering
    \begin{tabular}{|l|c|c|c|c|c|}
\hline
&\textbf{5}&\textbf{10}&\textbf{20}&\textbf{50}&\textbf{100}\\\hline
\textbf{1982-1995}&0.0149&0.0000&0.0000&0.0000&0.0000\\\hline
\textbf{1996-2001}&0.5564&0.1260&0.0073&0.0000&0.0000\\\hline
\textbf{2002-2006}&0.9992&0.9364&0.6062&0.1072&0.0058\\\hline
\textbf{2007-2013}&0.6657&0.1964&0.0185&0.0000&0.0000\\\hline
\textbf{2014-2017}&0.7592&0.2932&0.0426&0.0001&0.0000\\\hline
\textbf{2018-2021}&1.0000&0.9920&0.8394&0.2905&0.0428\\\hline
\end{tabular}
    \caption{p-values of Kolmogorov-Smirnov test statistic for the equality of the distributions of the social cost of carbon for the period subsample and the whole sample (rows) for different discretizations of the distribution (columns), for a 1\% pure rate of time preference.}
    \label{tab:ks1}
\end{table}

\begin{table}[hp]
    \centering
    \begin{tabular}{|l|c|c|c|c|c|}
\hline
&\textbf{5}&\textbf{10}&\textbf{20}&\textbf{50}&\textbf{100}\\\hline
\textbf{1982-1995}&0.6039&0.2070&0.0194&0.0000&0.0000\\\hline
\textbf{1996-2001}&0.6617&0.2386&0.0273&0.0000&0.0000\\\hline
\textbf{2002-2006}&1.0000&0.9780&0.7708&0.2106&0.0174\\\hline
\textbf{2007-2013}&0.9003&0.4477&0.0941&0.0009&0.0000\\\hline
\textbf{2014-2017}&1.0000&0.9780&0.7633&0.2074&0.0174\\\hline
\textbf{2018-2021}&1.0000&1.0000&1.0000&0.9999&0.9770\\\hline
\end{tabular}
    \caption{p-values of Kolmogorov-Smirnov test statistic for the equality of the distributions of the social cost of carbon for the period subsample and the whole sample (rows) for different discretizations of the distribution (columns), for a 2\% pure rate of time preference.}
    \label{tab:ks2}
\end{table}

\begin{table}[hp]
    \centering
    \begin{tabular}{|l|c|c|c|c|c|}
\hline
&\textbf{5}&\textbf{10}&\textbf{20}&\textbf{50}&\textbf{100}\\\hline
\textbf{1982-1995}&1.0000&1.0000&0.9984&0.8127&0.3586\\\hline
\textbf{1996-2001}&0.9973&0.9020&0.5364&0.0781&0.0030\\\hline
\textbf{2002-2006}&0.9605&0.6855&0.2544&0.0117&0.0001\\\hline
\textbf{2007-2013}&1.0000&0.9996&0.9627&0.5522&0.1610\\\hline
\textbf{2014-2017}&0.9993&0.9464&0.6450&0.1282&0.0081\\\hline
\textbf{2018-2021}&1.0000&1.0000&0.9893&0.7067&0.2755\\\hline
\end{tabular}
    \caption{p-values of Kolmogorov-Smirnov test statistic for the equality of the distributions of the social cost of carbon for the period subsample and the whole sample (rows) for different discretizations of the distribution (columns), for a 3\% pure rate of time preference.}
    \label{tab:ks3}
\end{table}

\begin{table}[hp]
    \centering
    \begin{tabular}{lc | rrrr | rr}
        weight & \%ile & \multicolumn{2}{c}{PRTP} & \multicolumn{2}{c |}{year} & \multicolumn{2}{c}{year only}\\
        \hline
         paper & & -131*** & (27) & -6.56* & (3.49)  & -3.18 & (3.36) \\
         & 0.1 & -9.50** & (4.34) & 0.288 & (0.599) & 0.480 & (0.832) \\
         & 0.3 & -15.5** & (6.5) & 0.487 & (0.891) & 1.31** & (0.55) \\
         & 0.5 & -37.1*** & (10.4) & 0.117 & (1.440) & 1.53 & (1.54) \\
         & 0.7 & -71.9** & (33.4) & -0.108 & (4.616) & 0.311 & (4.091)\\
         & 0.9 & -131 & (120) & -5.78 & (16.50) & -0.303 & (13.131)\\
         author & & -135*** & (26) & -6.29* & (3.36) & -3.24 & (3.21)\\
         & 0.1 & -8.71** & (4.25) & 0.583 & (0.564) & 0.576 & (0.777)\\
         & 0.3 & -15.2** & (6.4) & 0.610 & (0.852) & 1.35** & (0.53) \\
         & 0.5 & -35.8*** & (9.3) & 0.358 & (1.246) & 1.60 & (1.31) \\
         & 0.7 & -69.7** & (30.5) & -0.102 & (4.046) & 0.625 & (3.754)\\
         & 0.9 & -150 & (116) & -2.76 & (15.50) & -0.311 & (10.179)\\
         quality & & -104*** & (16) & -2.70 & (2.02) & 0.03 & (1.86) \\
         & 0.1 & -7.56*** & (1.70) & 0.569** & (0.225) & 0.406 & (0.272)\\
         & 0.3 & -11.6*** & (2.05) & 0.911*** & (0.271) & 1.39*** & (0.17)\\
         & 0.5 & -33.1*** & (3.8) & 0.705 & (0.508)& 2.08*** & (0.52) \\
         & 0.7 & -61.1*** & (11.5) & 0.200 & (1.520) & 1.56 & (1.35) \\
         & 0.9 & -137*** & (31) & 0.935 & (4.101) & 3.25 & (2.66) \\
         \hline
    \end{tabular}
    \caption{Results of weighted least squares and weighted quantile regression of the social cost of carbon on the pure rate of time preference and the publication year (middle columns) or publication year only (right columns).\\
    \footnotesize Standard errors are reported in brackets. Coefficients marked with ***, ** or * are statistically significant at the 1\%, 5\% or 10\% level, respectively. The top row has results for the mean regression, the next five rows for the quantile regression for the indicated percentile.}
    \label{tab:regress}
\end{table}

\begin{table}[hp]
    \centering
    \begin{tabular}{|l|c|c|c|c|c|c|c|}
\hline
&\textbf{3.0}&\textbf{2.0}&\textbf{1.5}&\textbf{1.0}&\textbf{0.1}&\textbf{0.0}&\textbf{other}\\\hline
\textbf{Q1}&0.1389&0.0068&0.0376&0.0413&0.0053&0.0076&0.0526\\\hline
\textbf{Q2}&0.0509&0.0095&0.0513&0.0331&0.0062&0.0057&0.0497\\\hline
\textbf{Q3}&0.0051&0.0096&0.0457&0.0322&0.0069&0.0070&0.0554\\\hline
\textbf{Q4}&0.0001&0.0103&0.0376&0.0294&0.0076&0.0095&0.0647\\\hline
\textbf{Q5}&0.0000&0.0144&0.0222&0.0255&0.0087&0.0214&0.0903\\\hline
\textbf{Null}&0.0390&0.0101&0.0389&0.0323&0.0070&0.0102&0.0625\\\hline
\end{tabular}
    \caption{Observed and hypothesized contribution to the kernel density of the social cost of carbon by quintile and pure rate of time preference.}
    \label{tab:discount}
\end{table}

\begin{table}[hp]
    \centering
    \begin{tabular}{|l|c|c|c|c|c|}
\hline
&\textbf{Hope}&\textbf{Nordhaus}&\textbf{Ploeg}&\textbf{Tol}&\textbf{Other}\\\hline
\textbf{Q1}&0.0185&0.0270&0.0205&0.0740&0.0882\\\hline
\textbf{Q2}&0.0234&0.0261&0.0276&0.0405&0.1063\\\hline
\textbf{Q3}&0.0212&0.0036&0.0247&0.0297&0.1120\\\hline
\textbf{Q4}&0.0185&0.0000&0.0216&0.0163&0.1201\\\hline
\textbf{Q5}&0.0094&0.0000&0.0176&0.0048&0.1485\\\hline
\textbf{Null}&0.0182&0.0114&0.0224&0.0331&0.1150\\\hline
\end{tabular}
    \caption{Observed and hypothesized contribution to the kernel density of the social cost of carbon by quintile and author.}
    \label{tab:author}
\end{table}

%figures

\begin{figure}[H]
    \centering
    \includegraphics[width=\textwidth]{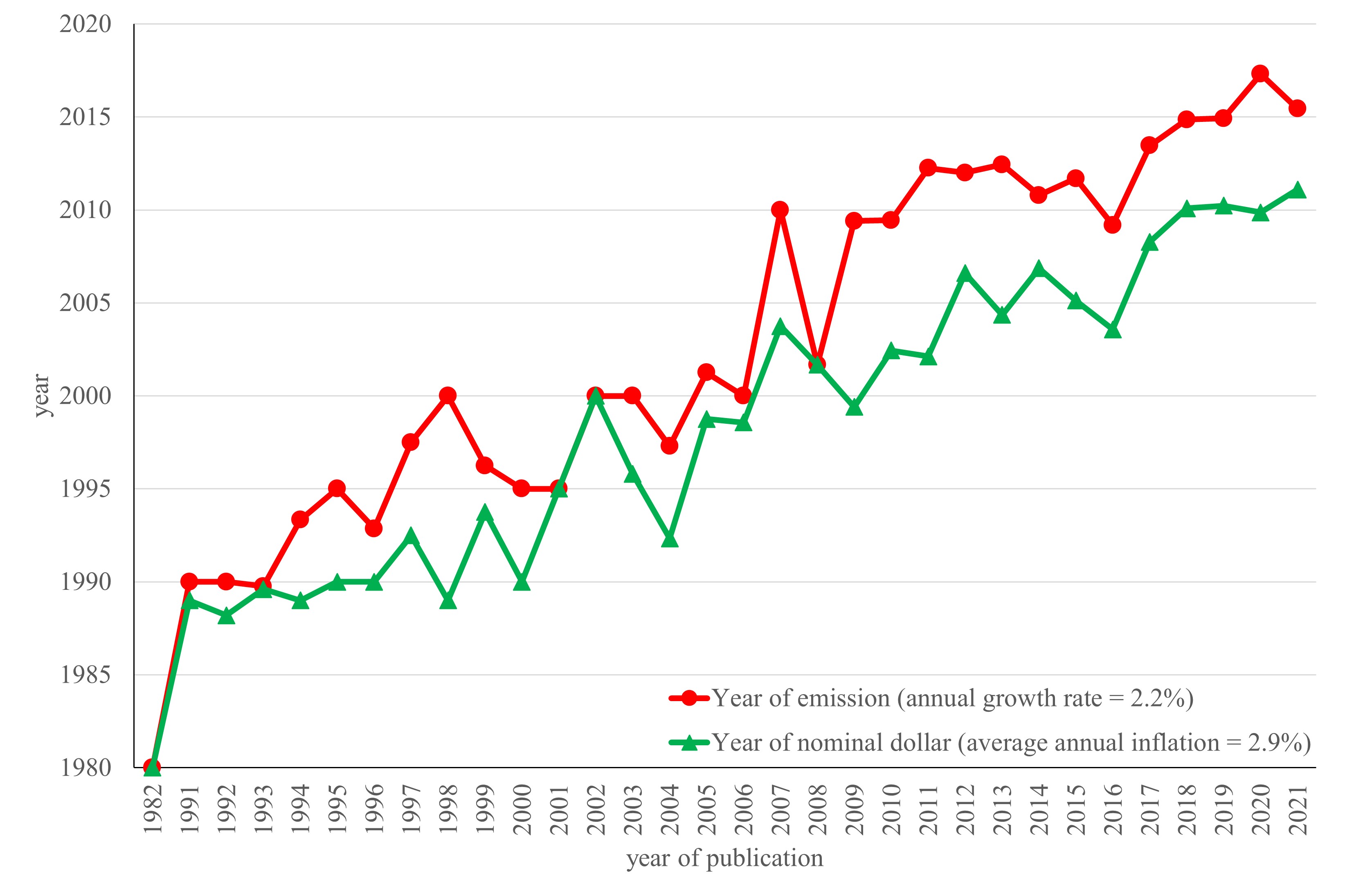}
    \caption{Year of emission and year of nominal dollar by year of publication. Estimates are weighted such that every published paper counts equally.}
    \label{fig:year}
\end{figure}

\newpage \printbibliography[title=Additional references]
\end{refsection}
%\end{comment}
\end{document}